\documentclass[pdflatex,sn-nature]{sn-jnl}

\usepackage{graphicx}%
\usepackage{multirow}%
\usepackage{amsmath,amssymb,amsfonts}%
\usepackage{amsthm}%
\usepackage{mathrsfs}%
\usepackage[title]{appendix}%
\usepackage{xcolor}%
\usepackage{textcomp}%
\usepackage{manyfoot}%
\usepackage{booktabs}%
\usepackage{algorithm}%
\usepackage{algorithmicx}%
\usepackage{algpseudocode}%
\usepackage{listings}%

\usepackage{bm}
\usepackage[varg]{txfonts}
\usepackage{enumerate}
\usepackage[normalem]{ulem}
\usepackage[OT1]{fontenc}
\usepackage{float} 
\usepackage{placeins}
\usepackage{diagbox}
\usepackage{scalerel,stackengine}

\providecommand{\citet}[1]{\citep{#1}}
\renewcommand{\citet}[1]{\citep{#1}}

\theoremstyle{thmstyleone}%
\theoremstyle{thmstyletwo}%

\theoremstyle{thmstylethree}%
\newcommand{\se}[1]{Section \ref{sec:#1}}

\newcommand{\Fig}[1]{Fig.~\ref{fig:#1}}

\newcommand{\be}{\begin{equation}}
\newcommand{\ee}{\end{equation}}
\newcommand{\bad}{\begin{equation} \begin{aligned}}
\newcommand{\ead}{\end{aligned} \end{equation}}
\newcommand{\Msun}{\,{M_\odot}}
\newcommand{\Mpc}{\,{\rm Mpc}}

\newcommand{\pc}{\,{\rm pc}}

\newcommand{\Myr}{\,{\rm Myr}}
\newcommand{\cm}{\,{\rm cm}}
\newcommand{\g}{\,{\rm g}}
\newcommand{\kms}{\,{\rm km/s}}

\newcommand{\rhos}{\rho_{\rm s}}

\newcommand{\Mv}{M_{\rm vir}}

\newcommand{\Mbh}{M_{\rm BH}}

\begin{document}

\title[DM origin of LRDs]{A Dark-matter Origin of Little Red Dots: Early Seeding and Super-Bondi Accretion}

\author[1]{\fnm{Hua-Peng} \sur{Gu}}

\author*[2,1]{\fnm{Fangzhou} \sur{Jiang}}\email{fangzhou.jiang@pku.edu.cn}

\author*[1,2]{\fnm{Xian} \sur{Chen}}\email{xian.chen@pku.edu.cn}

\author[3,4]{\fnm{Ran} \sur{Li}}

\author[1]{\fnm{Zi-Xiang} \sur{Jia}}

\affil[1]{\small\orgdiv{Department of Astronomy, School of Physics}, \orgname{Peking University}, \orgaddress{\city{Beijing}, \postcode{100871}, \country{China}}}

\affil[2]{\small\orgdiv{Kavli Institute for Astronomy and Astrophysics}, \orgname{Peking University}, \orgaddress{\city{Beijing}, \postcode{100871}, \country{China}}}

\affil[3]{\small\orgdiv{School of Physics and Astronomy}, \orgname{Beijing Normal University}, \orgaddress{\city{Beijing}, \postcode{100875}, \country{China}}}

\affil[4]{\small\orgdiv{School of Astronomy and Space Science}, \orgname{University of Chinese Academy of Sciences}, \orgaddress{\city{Beijing}, \postcode{100049}, \country{China}}}

\abstract{
The ``Little red dots” (LRDs) are a population of accreting supermassive black holes (SMBHs) in the early Universe which often exhibit undermassive or even undetectable stellar hosts\citep{Matthee+24,Chen+25,Jones+25}.  
Their early emergence, high space density, and extremely large black-hole-to-stellar mass ratios pose a serious challenge to conventional seeding scenarios that rely on baryon for both the formation and growth of black holes.  
Here we demonstrate that the above anomalies can be naturally resolved if dark matter is self-interacting \citep{Spergel+00,Kaplinghat+16}. 
We apply a fully relativistic, non-equilibrium halo-evolution model, first developed in our earlier work \citep{Gu+26a}, to trace the complete gravothermal evolution of self-interacting dark matter (SIDM) halos, from the initial collapse into BH seeds to the subsequent accretion of dark matter.  
We find that in highly concentrated halos assembled before reionization, gravothermal collapse efficiently produces stellar-mass black-hole seeds within a few hundred million years. 
Remarkably, and contrary to standard expectations for dark-matter accretion, heat conduction in SIDM then sustains a prolonged super-Bondi inflow that drives these seeds to supermassive scale by the LRD epoch, without baryonic assistance.  
The halo conditions required for completing these processes, together with the probability of avoiding major mergers that disrupt gravothermal evolution, result in an SMBH population consistent with the observed abundance and redshift distribution of LRDs.  
Our findings establish a pathway in which SMBHs are seeded and assembled primarily from dark matter, well before substantial galaxies form around them, thereby offering both a compelling physical explanation for LRDs and a 
new observational probe of dark-matter microphysics.
}

\maketitle

\section{Introduction}\label{sec:introduction}

The ``little red dots'' (LRDs) are a population of compact, unresolved sources
discovered by the James Webb Space Telescope in the early Universe that are
widely interpreted as accreting supermassive black holes (SMBHs)
\citep{Matthee+24,Kokorev+24,Greene+24,Maiolino+24}.  They are remarkably
abundant at high redshift, with inferred comoving number
densities of order $\sim 10^{-4}
\Mpc^{-3}$ at $4 \lesssim z \lesssim 9$, which could account for up to 1--10 per cent
of the galaxy population at that epoch
\citep{Kokorev+24,Greene+24,Perez-Gonzales+24}.  Many LRDs show only marginal
or undetectable stellar components, implying mass ratios between black holes
(BHs) and stellar hosts much higher than nearby values \citep{Maiolino+24}, or even
approaching unity in some systems \citep{Chen+25,Jones+25,Juodz+26}.  Their
early emergence, high abundance, and extreme BH-to-stellar mass ratios pose a
challenge to conventional baryonic SMBH-seeding scenarios within the standard
cosmological paradigm, including Population III remnants \citep{Chantavat+23},
very massive stars \citep{Kritos+24}, and direct-collapse black holes
\citep{Bromm+03,Begelman+06}.  
Population III remnants ($\sim10$--$100\Msun$) require sustained Eddington accretion despite strong stellar and BH feedback,
whereas the heavier seeds from runaway stellar mergers or direct gas collapse
($\sim10^{4}$--$10^{5}\Msun$) form only under rare conditions, including
extreme stellar densities, low metallicities, suppressed fragmentation, and
rapid gas inflow
\citep{Volonteri+10,Barack+19,Chantavat+23,Latif+13,Latif+16,Roberts+25,Inayoshi+20}.
These physical conditions are difficult to satisfy 
given the rapid star formation and chemical
enrichment inferred at the LRD epochs \citep{Dekel+23,Weibel+24}.
The unusual demographics of LRDs therefore motivate SMBH-formation pathways
that are less tightly coupled to stellar evolution and other baryonic
processes.

\begin{figure}[!b]
\centering
\includegraphics[width=0.95\linewidth]{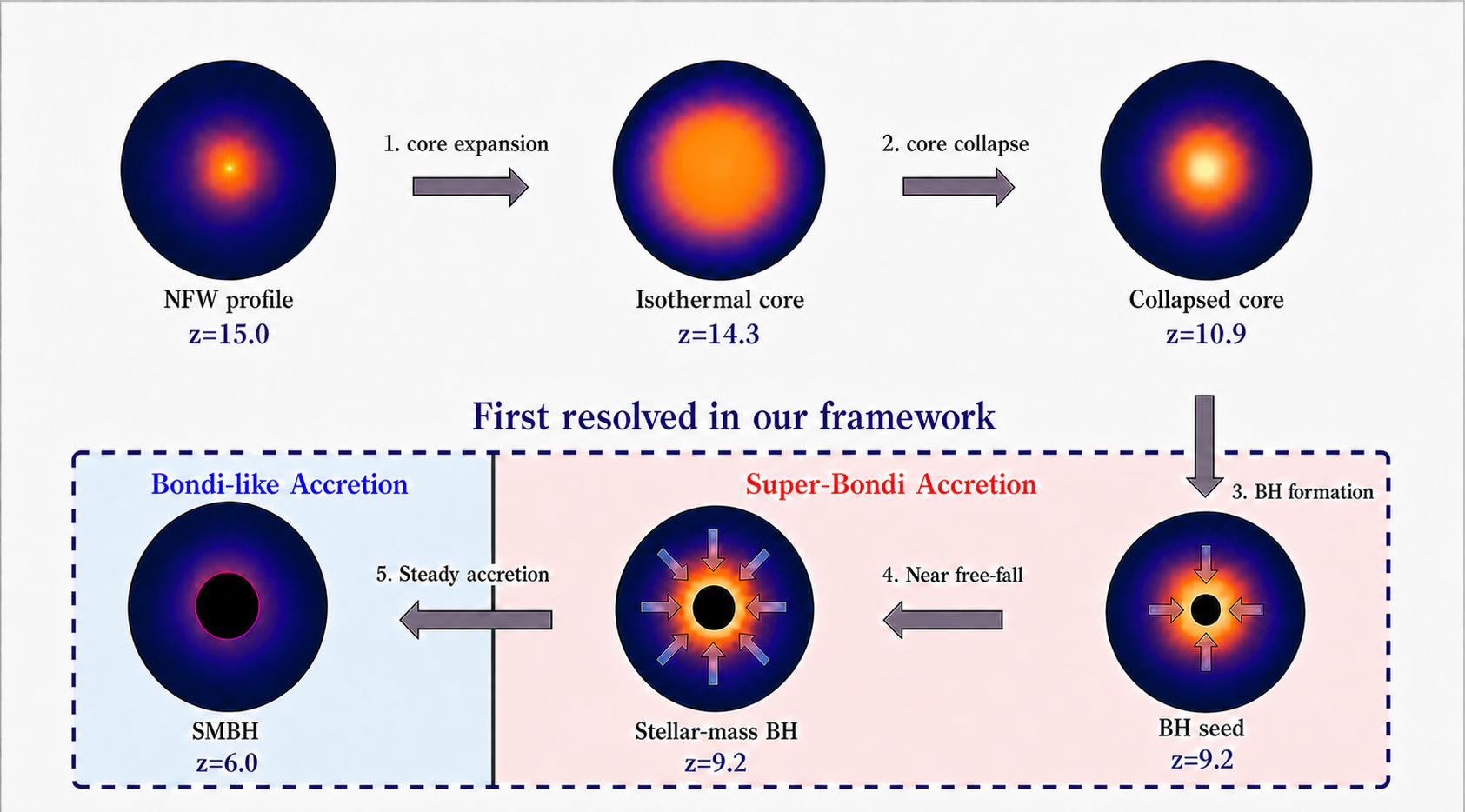}
\caption{
{\bf Five stages to form an SMBH out of an isolated SIDM halo.}
Our fiducial case is a halo with mass $\Mv=10^{7.5}\Msun$, formed at a pre-reionization epoch ($z=15$) with a Navarro--Frenk--White (NFW) profile of concentration $c=10$.
Self-scattering thermalizes the central part of the halo (Stage 1), which later
collapses (Stage 2) and produces a  
seed BH by $z=9.2$ (Stage 3). The seed BH subsequently grows by accreting the surrounding SIDM
at a super-Bondi rate, due to the 
near-free-fall SIDM inflow (Stage 4) and heat conduction effects (early Stage 5)
until
the BH reaches the supermassive scale by $z\sim6$.
The late stages of gravothermal evolution (Stages 3--5) are physically resolved for the first time 
in this work.
}
\label{fig:illustration}
\end{figure}

Dark matter provides one such possibility. Among alternative dark-matter
models, self-interacting dark matter (SIDM) has attracted broad interest across galaxy
formation \citep{Spergel+00,Vogelsberger+12,Rocha+13,Kaplinghat+16}, particle
physics \citep{Feng+09,Loeb+11,Tulin+13}, and compact-object astrophysics
\citep{Pollack+15,Choquette+19}. By introducing elastic self-scattering
between dark-matter particles, SIDM, while preserving the large-scale successes of the
cold dark-matter paradigm \citep{Rocha+13}, can address a broad range of small-scale
phenomena with only two additional parameters specifying a velocity-dependent
self-scattering cross section \citep{Loeb+11,Tulin+13}.  In dense halo centers,
self-scattering enables heat transport and drives a multi-stage gravothermal
evolution (\Fig{illustration}).  The inner halo first develops a nearly
isothermal, low-density core, potentially accounting for the shallow central
density profiles observed in many galaxies
\citep{Tulin+18,Adhikari+22,Zhang+25}.  Continued outward heat conduction,
together with the negative heat capacity of self-gravitating systems, can
subsequently reverse this evolution and trigger runaway gravothermal core
collapse \citep{Balberg+02a,Koda+11,Pollack+15,Feng+22,Outmezguine+23}.  These different
evolutionary stages allow SIDM to produce a diverse range of subgalactic
structures within a unified framework, from cored, low-surface-brightness
galaxies \citep{Zhang+25} to the highly concentrated halos inferred from
gravitational lensing \citep{Minor+21,Yang+21,Yu+26,Li+25,Cao+26}.  Taken to
its extreme, gravothermal collapse can drive the central SIDM density
sufficiently high for BH  formation
\citep{Balberg+02a,Koda+11,Pollack+15,Feng+21,Feng+22,Gu+26a}, opening a pathway in
which BH seeds originate directly from dark-matter halos.

Gravothermal core collapse of SIDM halos has therefore been proposed as a
pathway to form high-redshift SMBHs associated with LRDs.  Previous work showed
that this scenario can reproduce the observationally inferred BH mass function
of LRDs \citep{Jiang+25} using self-scattering cross sections consistent with
independent constraints from nearby-galaxy rotation curves \citep{Jia+26}.
However, these predictions assumed that a fixed fraction ($\sim 1\%$) of the
collapsing halo forms the BH seed \citep{Pollack+15,Feng+21}.  Determining the
seed mass from first principles requires following gravothermal collapse into
its highly nonlinear and relativistic regime, beyond the reach of
quasi-equilibrium, non-relativistic fluid treatments
\citep{Shapiro+18,Jiang+23,Gad-Nasr+24} or current $N$-body simulations
\citep{Palubski+24,Fischer+24}.  Existing calculations therefore terminate
during the early stages of gravothermal evolution (Stages 1 and 2 in
\Fig{illustration}), before the formation of an apparent horizon (Stage 3) and
the subsequent dark-accretion phases (Stages 4--5).  The mass of the resulting
BH seed, and, equally importantly, the density and dynamical state of the SIDM
remaining around it, has consequently remained uncertain.

An equally important open question is whether a BH seed formed through SIDM
core collapse can subsequently grow to supermassive scales within the limited
cosmic time available.  Previous studies have considered dark-matter accretion onto BH seeds artificially embedded in SIDM halos \citep{Hu+06a,Sabarish+25,Meng+26},
generally finding growth too slow to assemble SMBHs by $z\sim6$ without
additional baryonic effects \citep{Lora-Clavijo+14,Feng+25,Frank+25}.  However, a BH born from gravothermal collapse is not in an equilibrium halo: it emerges from a rapidly evolving, highly concentrated SIDM core whose thermodynamic and dynamical states are inherited directly from the collapse itself.  
Assessing its subsequent growth therefore requires following seed formation and dark-matter accretion as a single, continuous process. 
Furthermore, heat conduction in SIDM may significantly reduce central thermal support and facilitate accretion, an effect neglected in previous studies. 
To quantify this effect, we extend our relativistic, non-equilibrium Misner–Sharp framework \citep{Gu+26a} (Methods) to evolve SIDM halos continuously from gravothermal collapse through horizon formation and into the ensuing accretion phase (\Fig{illustration}). We apply this method to pre-reionization halos over a range of masses, concentrations and self-interaction parameters, and combine these calculations with cosmological halo abundances and merger histories to investigate whether the resulting conductive enhancement can make dark-matter accretion alone sufficiently rapid to account for the SMBHs of LRDs. 

\FloatBarrier
\section{Dark seeding and dark accretion}\label{sec:results}

We first illustrate the SIDM pathway using a fiducial pre-reionization halo before exploring how SMBH formation depends on halo properties. 
This example follows the complete evolution from gravothermal collapse to BH seeding and subsequent dark-matter accretion, providing the physical basis for the parameter study presented in \se{discussion}.

Our fiducial halo has a virial mass of $\Mv=10^{7.5}\Msun$ and forms at $z=15$, when sufficiently massive halos already exist but baryonic effects remain relatively weak.\footnote{At $z=15$, the adopted halo mass corresponds to a $\sim3\sigma$ density peak and lies below the atomic-cooling threshold \citep{Barkana+01,Reed+07}.} 
We initialize the halo with a Navarro--Frenk--White (NFW) density profile \citep{Navarro+96} and a concentration of $c=10$ (corresponding to scale radius $R_\mathrm{s}=60\pc$), a relatively high but realistic value in cosmological simulations \citep{Jiang+25}. 
We adopt a velocity-dependent SIDM model with $\sigma_0=20\,\mathrm{cm^2\,g^{-1}}$ and $w=100\kms$, consistent with independent constraints from nearby galaxy rotation curves \citep{Jia+26} (Methods). 
We later vary these parameters to determine the halo conditions required for timely SMBH formation (\se{discussion}).

\subsection{BH seed formation}\label{sec:seeding}

\begin{figure*}[!t]
\centering

\makebox[\linewidth][c]{%
    \hspace*{-0.3cm}%
    \includegraphics[width=0.75\linewidth]{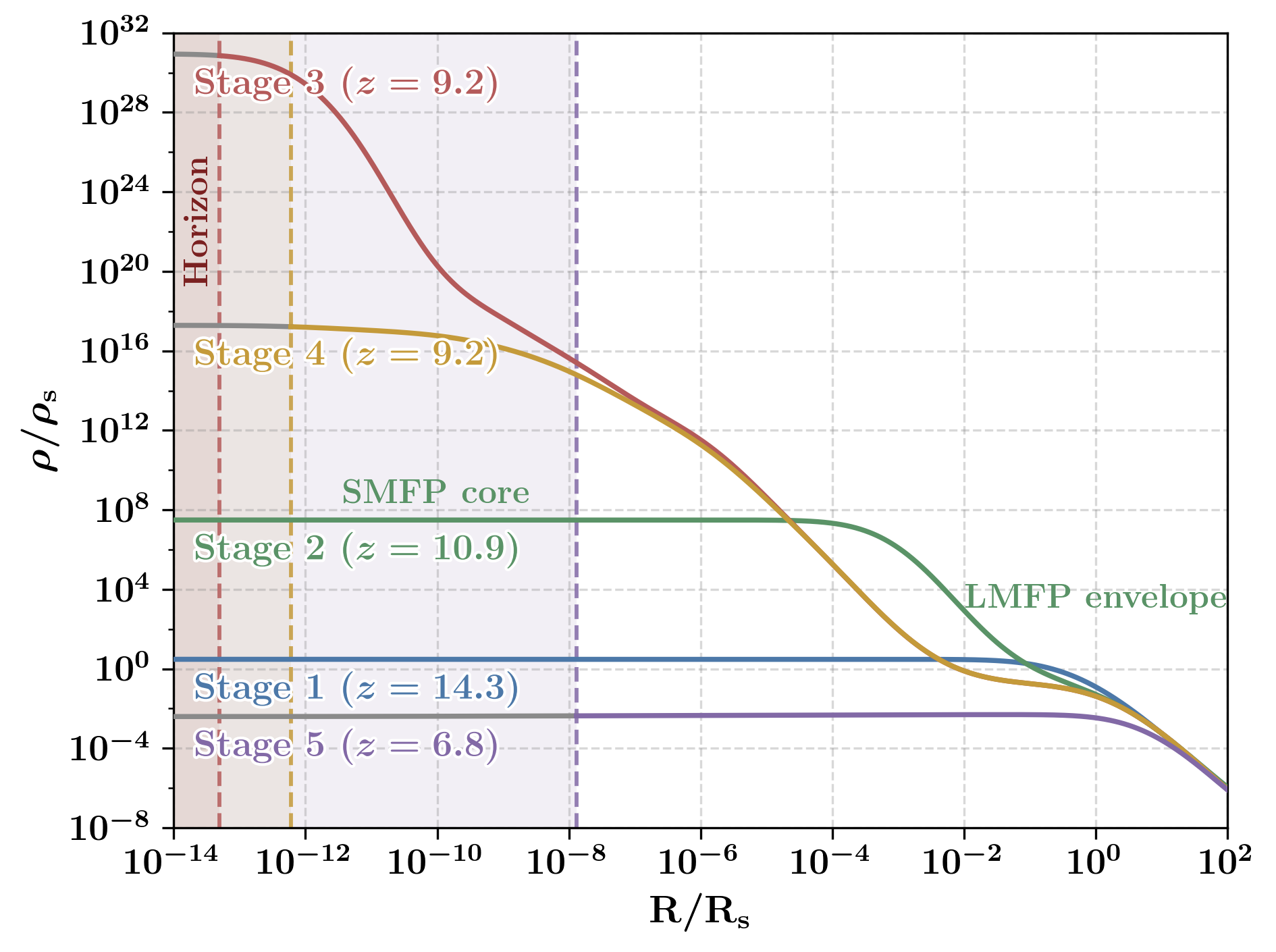}%
    \hspace*{0.3cm}%
}\par

\includegraphics[width=0.77\textwidth]{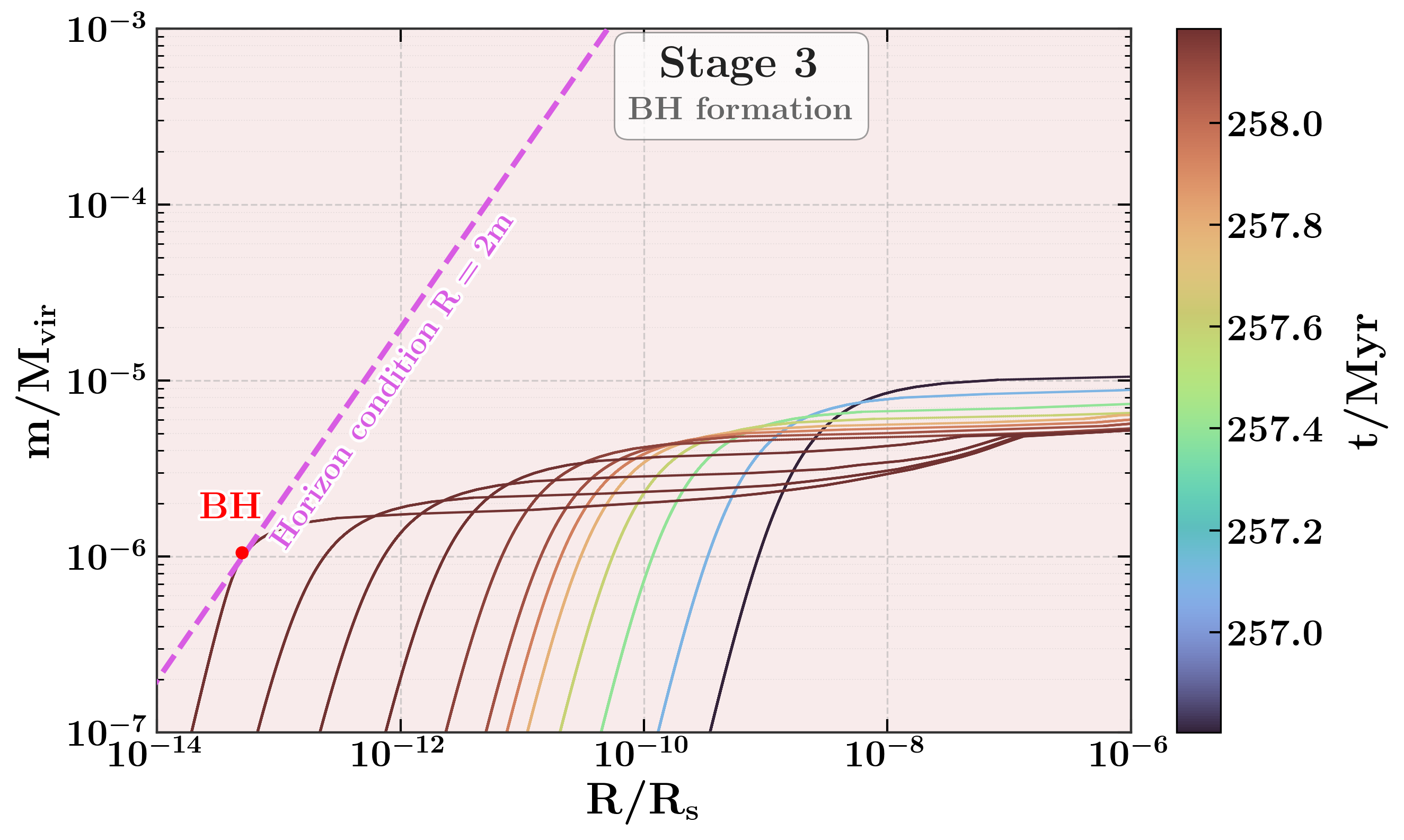}

\caption{
{\bf Evolution of SIDM halo profile, illustrated by the fiducial experiment.} 
\quad
{\it Upper}: Five representative density profiles corresponding to the evolutionary stages illustrated in \Fig{illustration}. During gravothermal collapse (Stage 2), a dense central core develops and contracts until it forms a BH seed (Stage 3). The seed then grows by accreting the remaining core (Stage 4) and then as well as the surrounding envelope (Stage 5). The horizon radii in Stages 3--5 are indicated by the vertical dashed lines of corresponding colors.
\quad
{\it Lower}: A zoomed-in view of the mass profiles immediately before the onset of BH formation (Stage 3), showing the first appearance of an apparent horizon and the emergence of a $\sim 30\Msun$ BH seed at $t=258\Myr$. The color scale denotes the time elapsed since halo formation at $z=15$.
}
\label{fig:horizon}
\end{figure*}

Schematically illustrated in \Fig{illustration},
heat conduction drives the fiducial SIDM halo through two well-known stages of gravothermal evolution: initial core expansion (Stage 1), followed by runaway core collapse (Stage 2) \citep{Gu+26a}. 
During the late stages of collapse, a dense central core develops within a lower-density envelope by $t\sim150\Myr$ ($z=10.9$; \Fig{horizon}).\footnote{Here $t$ denotes the elapsed time since halo formation at $z=15$. }
The core then contracts while continuously losing mass until reaching the threshold for BH formation.
Owing to the velocity-dependent self-interaction cross section, the late-stage evolution deviates from the standard self-similar gravothermal-collapse solution \citep{Balberg+02a}.

At $t=258\Myr$ ($z=9.2$), the mass profile first satisfies the apparent-horizon condition $R=2m$ in geometrized units ($G=c=1$), where $R$ is the areal radius and $m$ the enclosed Misner--Sharp mass (\Fig{horizon}, lower panel).
This occurs at $m=10^{-6}\Mv$, signaling the formation of a central BH seed with a mass of $\sim30\Msun$. 
This rapid collapse is enabled by two factors: the high characteristic densities of early-Universe halos, which substantially shorten the SIDM relaxation timescale relative to their lower-redshift counterparts (e.g., see \cite{Gu+26a} for comparison), and the velocity-dependent self-interaction cross section, which accelerates the final collapse phase (Stage 2, see Methods; \Fig{cross_section}).
Although the resulting seed is relatively light, SIDM offers an important demographic advantage: BH seeds arise naturally in sufficiently concentrated halos, circumventing the need for rare baryonic environments invoked by conventional seeding models.

\subsection{Super-Bondi dark accretion} \label{sec:accretion}

\begin{figure}[!htbp]
\centering
\includegraphics[width=0.77\linewidth]{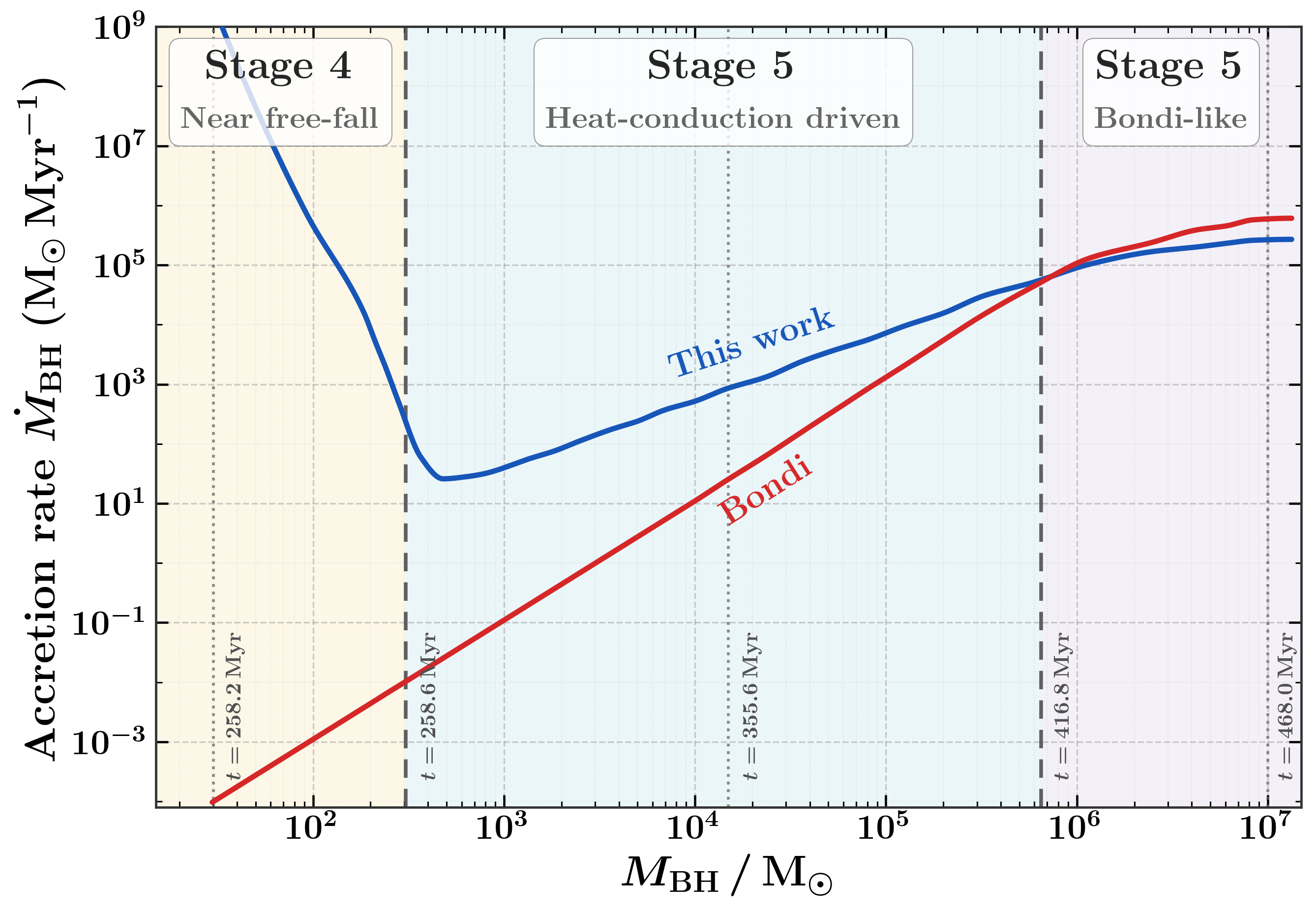}
\caption{
{\bf Comparison of the accretion rate in our fiducial model (blue) and the Bondi prediction (red).}
The Bondi rate scales as $\dot{M}_{\rm B}\propto \Mbh^{2}$, whereas the dark
accretion proceeds through three regimes: near-free-fall inflow during Stage 4,
heat-conduction-driven super-Bondi accretion during most of Stage 5, and
Bondi-like accretion at late times when BH gravity dominates the inflow.
} \label{fig:accretion} \end{figure}

\begin{figure}[!htbp]
\centering
\includegraphics[width=0.77\linewidth]{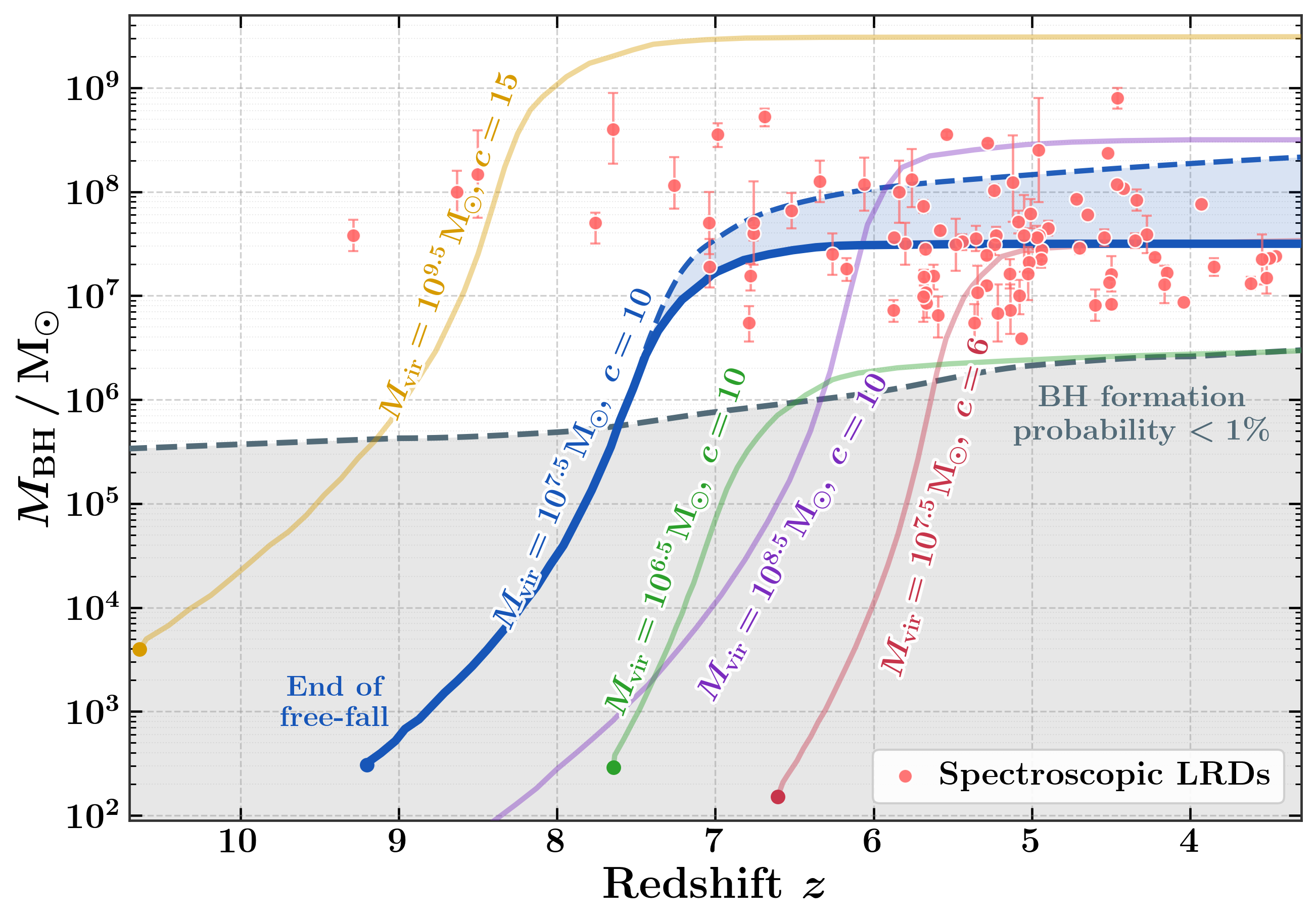}
\caption{
{\bf Growth of the BHs seeded by core-collapsed SIDM halos,
and a comparison with the observed LRDs. } 
The blue solid curve is the predicted BH mass–redshift tracks our fiducial model, in which the BH seed forms at $z\sim9$, subsequently accretes nearly the entire SIDM halo and
reaches $\Mbh\sim10^{7.5}\Msun$ by $z\sim6$, characteristic of observed LRDs (red data points, from
\citep{Ivey+26,Jones+25,Kocevski+24,Loiacono+26,Maiolino+26,Naidu+25,Taylor+25,Zhuang+25}
-- only LRDs with spectroscopic BH-mass measurements are included).
Varying the initial halo mass and concentration (but keeping halo formation at the pre-reionization epoch of $z=15$) produces growth tracks spanning
the observed LRD distribution (other solid lines). 
SIDM core collapse would be interrupted by major mergers (grey dashed line; see Methods and \Fig{merger}), naturally defining a redshift-dependent characteristic mass below which SMBH formation through this mechanism is unlikely. 
Intriguingly, this scale coincides with the uncertain observational lower bound on BH masses of $\sim10^{5-6}\Msun$ \citep{Greene+26}.
In a cosmological environment, smooth halo accretion can continuously replenish
the SIDM reservoir, allowing further BH growth (blue dashed line). This smooth-accretion limit also sets a lower bound on the LRD host-halo mass by the LRD epoch.
} \label{fig:observation} \end{figure}

The newly formed BH seed is embedded in a dense residual SIDM core, providing
ideal conditions for subsequent dark accretion.  Immediately after formation,
the BH seed is embedded within a dark-matter core with density $\rho\sim10^{30}\rhos$,
surrounded by a much lower-density envelope with $\rho\sim0.1\rhos$, where
$\rho_s$ denotes the scale density of the initial NFW halo (\Fig{horizon}, upper panel).
The BH first rapidly accretes the residual core (Stage 4), after which the
outer envelope continuously replenishes the central region, sustaining a
prolonged phase of steady accretion (Stage 5). \Fig{accretion} compares the resulting accretion rate onto the central BH with
the corresponding Bondi rate.  The Bondi solution assumes a quasi-static
ambient medium, in which the central BH gravity drives a perturbative inflow.
At the onset of Stage 4 ($t=258.2\Myr$), however, the residual core lies
immediately outside the newly formed horizon.  Here, strong
general-relativistic effects can overcome thermal pressure support
\citep{Gu+26a} to trigger a catastrophic inflow into the BH (Methods).  The BH
mass therefore rises rapidly from $\sim30\Msun$ to $\sim300\Msun$ within only
$0.4$ Myr.  This previously unexplored accretion phase is resolved
self-consistently in our extended Misner--Sharp framework.

As the residual core is depleted, the BH accretion rate reaches a local minimum
(\Fig{accretion}) yet remains highly super-Bondi.  The BH then begins to
accrete the outer dark-matter envelope, marking the final stage of gravothermal
evolution (Stage 5).  Throughout most of this stage, the BH continues to
accrete at a rate approximately two orders of magnitude above the Bondi rate,
consistent with the independent findings of \citet{Sabarish+25}.  This
enhancement is driven by SIDM heat conduction, which continuously transports
energy outward, weakens central thermal pressure support, and sustains an
inward mass flux.  A quantitative interpretation of this super-Bondi phase is
provided in Methods. As the BH grows toward supermassive scales, gravitational
capture gradually overtakes gravothermal inflow, causing the accretion rate to
asymptotically approach the Bondi rate, as shown in \Fig{accretion}. 

\Fig{observation} shows the corresponding evolution of the BH
mass.  The growth track from our fiducial model (blue line) passes right through
the region occupied by observed LRDs (red points), at $4\lesssim z\lesssim7$
and $10^{6.5}\Msun\lesssim \Mbh \lesssim 10^9\Msun$.  Varying the initial halo
properties within plausible ranges produces growth tracks that collectively
span the observed LRD distribution in the $\Mbh$--$z$ plane. 
Another interesting aspect is that halos with $\Mv\lesssim10^6\Msun$ undergo frequent mergers at the LRD epochs (Methods; \Fig{merger}), interrupting gravothermal evolution and thereby suppressing the formation of lower-mass SMBHs through the SIDM channel (grey dashed line in \Fig{observation}). Intriguingly, the resulting characteristic BH mass coincides with the lower bound of observationally inferred LRD BH masses, although this bound remains uncertain \citep{Greene+26}.
In our fiducial isolated-halo calculations, BH growth ceases once
the original SIDM halo is largely consumed (blue solid line in \Fig{observation}), reaching
$\Mbh\sim10^{7.5}\Msun$ by $t\approx500\Myr$ ($z\sim6$). 
This result should be interpreted with caution and, in particular, does not imply that the seed BH necessarily consumes the entire SIDM halo. 
The fluid approximation breaks down once the BH has consumed approximately 20 \% of the halo, as the declining envelope density -- and hence the SIDM  scattering rate -- becomes too low for the fluid treatment to remain valid (Methods). 
Moreover, in a cosmological setting, ongoing smooth halo accretion can replenish the SIDM reservoir as it is consumed by the central BH, allowing further growth (blue dashed line in \Fig{observation}).  
The SIDM pathway therefore naturally produces SMBHs with the masses and redshifts characteristics of the observed LRD population.

\FloatBarrier
\section{Cosmological SMBH demographics and observational predictions}\label{sec:discussion}

\begin{figure*}[!ht]
\centering
\includegraphics[width=0.77\textwidth]{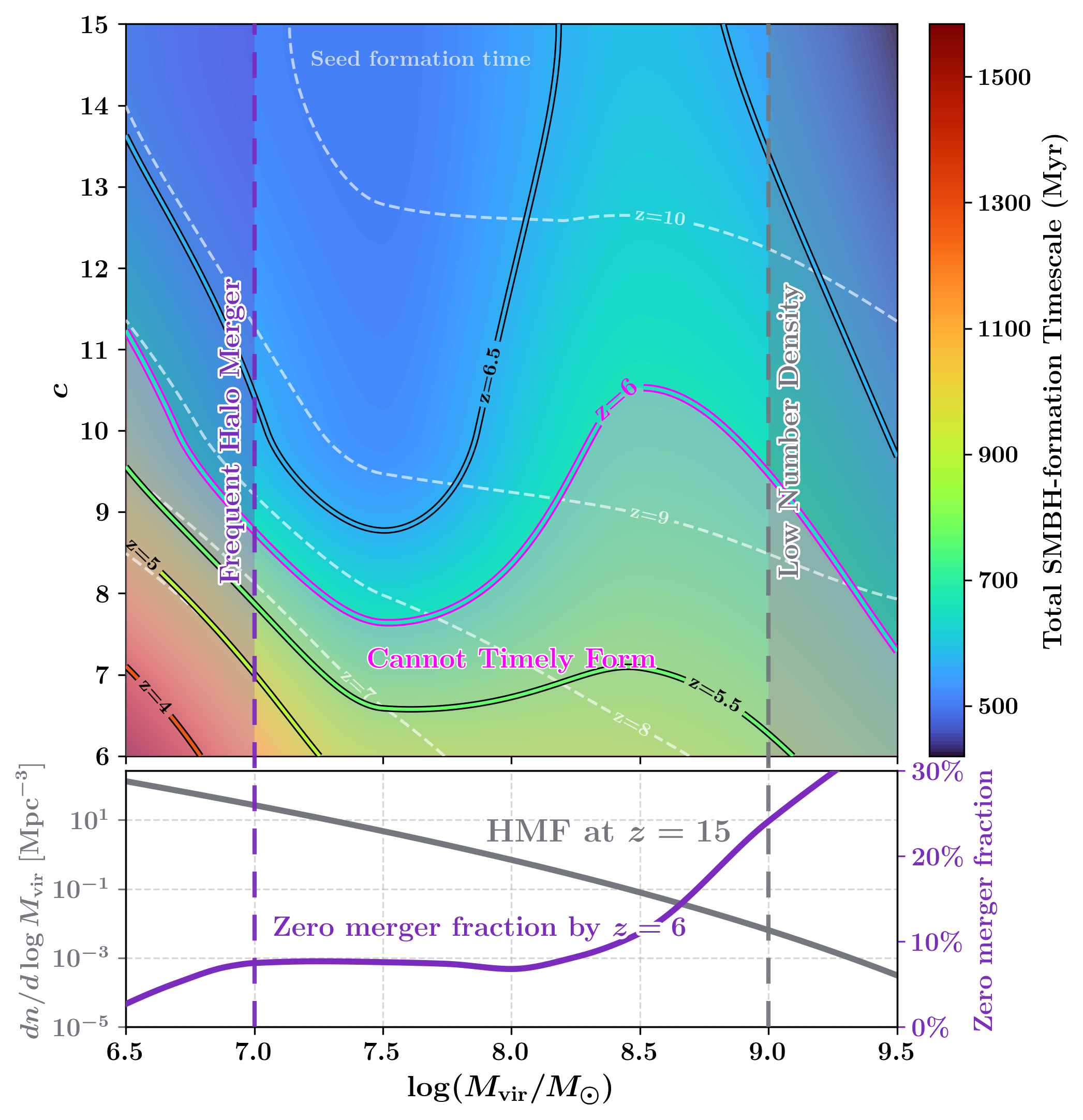}\par
\caption{
\textbf{Halo conditions for timely BH seeding and growth into SMBHs} -- dependence of SMBH formation time on halo virial mass, $\Mv$, and concentration, $c$, for a self-interaction cross section of $\sigma_0=20\cm^2 \g^{-1}$ and $w=100\kms$.
\quad
{\it Upper}: Time elapsed from halo formation at $z=15$ until the BH grows to supermassive scales by accreting half of the initial halo mass. 
In general, more massive and more concentrated halos evolve more rapidly. 
Halos with $\Mv\gtrsim10^7\Msun$ and $c\gtrsim10$ form SMBHs before $z=6$ (magenta line).  
BH-seed formation time under different halo conditions is denoted by the white dashed contours.
\quad
{\it Lower}:
Two additional factors governing the abundance of viable SMBH halos: the fraction of halos that avoid major mergers by $z=6$ (purple) and the halo mass function (HMF) at $z=15$ (grey). 
That is, halos with $\Mv\lesssim10^7\Msun$ are unlikely to produce SMBHs in time because gravothermal evolution is frequently interrupted by major mergers (purple dashed line), while very massive halos are intrinsically rare (grey dashed line).
Together with the collapse and growth times shown above, these trends naturally define a preferred range of initial halo masses capable of hosting LRDs.
}
\label{fig:scan}
\end{figure*}

\subsection{Host-halo conditions and LRD abundance}

We first identify the halo conditions required for SMBHs to form by the LRD
epoch using a parameter survey shown in \Fig{scan}.  For our fiducial SIDM
model, halos initialized at $z=15$ with $\Mv\gtrsim10^7\Msun$ and
concentrations $c\gtrsim10$ generally complete both gravothermal collapse and
the subsequent dark-accretion phase by $z\sim6$.  The detailed dependence of
the formation timescale on halo mass and concentration is discussed in Methods.
Here we adopt these criteria to predict the cosmological abundance of LRDs.
The predicted abundance follows directly from three independently determined
ingredients.  The comoving number density of halos with $\Mv>10^7\Msun$ at
$z=15$ is $7.7\Mpc^{-3}$ \citep{Murray+13}.  At these redshifts, halo
concentrations are nearly independent of mass, and approximately $0.11\%$ of
halos satisfy $c>10$ \citep{Jiang+25}.  Together, these conditions imply a
number density of $8\times10^{-3}\Mpc^{-3}$ for halos capable of producing
SMBHs through SIDM core collapse.

However, not every such viable halo is expected to successfully form an SMBH
and be detected as an LRD.  
Major mergers inject orbital kinetic energy into the halo center, resetting the gravothermal evolution and preventing core collapse \citep{Silverman+26}. 
Using cosmological merger trees (Methods; \Fig{merger}), we find that only $\gtrsim8\%$ of viable halos avoid a major merger throughout their evolution,  reducing the predicted SMBH number density to
$6\times10^{-4}\Mpc^{-3}$.  Additionally, the duty cycle of high-redshift
active galactic nuclei is expected to be 20-30 percent \citep{Roy+26,Shen+26}.
The resulting abundance is therefore in good agreement with the observed LRD
number density of $\sim10^{-4}\Mpc^{-3}$ \citep{Perez-Gonzales+24}.
Integrating over the halo mass function further gives a total SIDM-seeded SMBH
mass density of $\rho_{\rm BH}\approx8\times10^3\Msun\Mpc^{-3}$, about an order
of magnitude below the local SMBH mass density \citep{Shankar+04}, leaving
ample room for subsequent baryonic growth.  Together, these results show that
the SIDM scenario quantitatively reproduces the observed abundance of LRDs
while remaining consistent with local constraints.

\subsection{The characteristic redshift and halo-mass scales of LRDs}

Beyond the abundance comparison above, our model makes additional predictions for the redshift and mass distributions of SIDM-seeded BHs. Encouragingly, these trends are consistent with the properties of the current LRD sample.
Before $z\gtrsim9$, the cosmic time budget is simply too short for gravothermal collapse and subsequent dark accretion to
assemble SMBHs.  At lower redshifts, although suitable halos become
increasingly common, uninterrupted gravothermal evolution becomes progressively
rarer because hierarchical structure formation produces frequent major mergers.
By $z\sim4$, most halos have experienced at least one major merger
(\Fig{merger}), resetting the gravothermal clock and suppressing new
SMBH formation. 
The combination of these effects places the preferred epoch of
SIDM-seeded SMBH formation at $4\lesssim z\lesssim9$, coincident with the
redshift range occupied by the observed LRD population.

Merger statistics also select a characteristic halo-mass range for LRD hosts.
Halos with $\Mv\lesssim10^7\Msun$ rarely remain free of disruptive major mergers  (\Fig{scan}; \Fig{merger}), which can interrupt gravothermal evolution and prevent SMBH formation even when the nominal collapse timescale is sufficiently short \citep{Silverman+26}.
Combined with the gravothermal constraints shown in \Fig{scan}, 
formation of LRDs favors halos of $\Mv\sim10^7$–$10^9\Msun$ during the pre-reionization epoch.
By the time these systems are observed as LRDs, continued cosmological mass assembly can increase their masses by approximately $0.5$–$1$ dex \citep{Jiang+20}.
Their host halos nevertheless remain substantially less massive than those
inferred for luminous quasars \citep{Pizzati+25}, providing a testable
prediction through future measurements of LRD clustering and large-scale
environments.

\subsection{Observational signatures and tests of the SIDM scenario}

Because the SIDM scenario explicitly links SMBH formation to the properties and assembly histories of dark-matter halos, it makes several distinctive predictions that can be tested with observations.

First, LRDs should preferentially inhabit early-forming, highly concentrated halos with quiescent assembly histories. 
Their host halos should therefore exhibit characteristic assembly bias and spatial clustering. 
Cosmological simulations that combine SIDM halo evolution with realistic merger histories will enable quantitative predictions for these signatures.
Notably, LRD hosting halos grow into low-mass host halos of $\Mv\lesssim10^{9.5}\Msun$, which corresponds to a halo bias of $b\sim2$; yet, the requirement of early formation and high concentration yields assembly bias that increases the clustering strength to $b\sim3$ \citep{Wang+26}. 

Second, the SIDM pathway allows for the possible existence of nearly “naked” SMBHs. 
In the absence of disruptive mergers, a BH formed through SIDM core collapse can consume a substantial fraction of its original host halo (\Fig{observation}). 
Although such uninterrupted evolution is expected to be uncommon, it may already have observational support. 
A recent dynamical analysis of a gravitationally lensed LRD found gas motions consistent with an almost purely Keplerian potential, suggesting a system containing little stellar mass and possibly little remaining dark matter \citep{Juodz+26}.

Finally, the demographics of LRDs offer a new probe of SIDM microphysics. 
The self-interaction cross section influences both gravothermal collapse and the subsequent dark-accretion phase in opposite ways: reducing the cross section delays BH seed formation but also weakens collisional pressure support, enhancing late-time accretion. 
Over the range explored here, these competing effects largely compensate, leaving the predicted SMBH abundance relatively insensitive to the cross section around the values we adopted (Methods). 
Extending the calculations across a broader SIDM parameter space should identify the regions where either timely seed formation or subsequent growth becomes impossible. 
Joint measurements of the LRD abundance, redshift evolution, clustering, and BH-mass distribution could therefore provide complementary constraints on the dark-matter physics that are inaccessible through any single observable or nearby-galaxy kinematics alone.

\FloatBarrier
\section{Methods}\label{sec:methods}

\subsection*{Extended Misner--Sharp formalism}

The simulations in this study are based on the Misner--Sharp framework \citep{Misner+64}, which provides a general-relativistic, non-equilibrium description of fluid dynamics. In our previous work \citep{Gu+26a}, we first adapted this framework to model SIDM halo evolution by incorporating heat conduction in SIDM. Unlike conventional fluid treatments of SIDM halos, this extension allows us to capture BH seed formation and follow the ensuing accretion within a single, unified framework. 

More specifically, we adopt the time-dependent, spherically symmetric
metric
\begin{equation}\label{eq:Eq1}
ds^2 = -e^{2\phi(r,t)}dt^2 + e^{\lambda(r,t)} dr^2 + R(r,t)^2d\Omega^2,
\end{equation}
where $r$ is the comoving radial coordinate, $R(r,t)$ is the circumferential
radius, and $\phi(r,t)$ is the gravitational potential.

To model heat conduction in SIDM, we extend the perfect-fluid energy--momentum tensor to include the heat-flux four-vector $q^\mu$
\begin{equation}\label{eq:Eq2}
T^{\mu\nu} = [\rho(1+\epsilon) + P] u^\mu u^\nu + P g^{\mu\nu} + q^\mu u^\nu + q^\nu u^\mu,
\end{equation}
where $\rho$ is the rest-mass density, $\epsilon$ is the specific internal
energy, $P$ is the pressure, and $u^\mu$ is the fluid four-velocity. Assuming
sufficiently frequent SIDM collisions, we close the system with the ideal-gas
equation of state $P=(\gamma-1)\rho\epsilon$, where $\gamma$ is the adiabatic
index. Following Eckart's relativistic thermodynamics
\citep{Eckart+40,Shapiro+18}, we express the radial heat flux $q^r$ as
\begin{equation}\label{eq:Eq3}
q^r = -(\gamma-1)\kappa e^{-\lambda} e^{-\phi} \frac{\partial}{\partial r}\left(\epsilon e^{\phi}\right),
\end{equation}
where $\kappa$ is the thermal conductivity.

To capture nonequilibrium collapse, the formalism tracks the bulk radial velocity $U\equiv e^{-\phi}\partial R/\partial t$ by solving the acceleration equation
\begin{equation}\label{eq:Eq4}
e^{-\phi}\frac{\partial U}{\partial t} = \Gamma e^{-\lambda/2}\frac{\partial\phi}{\partial r} - \frac{m + 4 \pi R^3 P}{R^2},
\end{equation} 
where $m$ is the enclosed mass and $\Gamma=\sqrt{1-2m/R+U^2}$ is
the generalized Lorentz factor. 
Equation~(\ref{eq:Eq4}) explicitly couples
general-relativistic gravity to the pressure gradient. As the apparent horizon
is approached ($R\to2m$), $\Gamma\to0$. This relativistic effect strongly
suppresses the outward pressure-gradient term $\partial\phi /\partial r$ (see \citep{Gu+26a}), reducing thermal support and
triggering the rapid dynamical inflow during the initial accretion phase (Stage
4).

We resolve central BH formation by tracking outgoing radial null geodesics
\citep{Hernandez+66}. A trapped surface forms when $R=2m$, marking the apparent
horizon \citep{Penrose+65,Hawking+23}. This condition provides a
self-consistent criterion for determining the initial BH seed mass without
nonrelativistic approximations. A complete derivation of the coupled
hydrodynamic equations is provided in \citet{Gu+26a}.

\subsection*{Velocity-dependent cross section}

Compared with \citet{Gu+26a}, a key extension in this work is the adoption of a velocity-dependent cross section \citep{Gilman+21b,Yang+22,Zeng+25}
\begin{equation}\label{eq:Eq5}
\sigma(v) = \frac{\sigma_\mathrm{0}}{\left(1+(v/w)^2\right)^{2}},
\end{equation}
where $v=\sqrt{P/\rho}$ is the SIDM velocity dispersion. For the fiducial model, we adopt the observationally favored values $\sigma_0=20\,\mathrm{cm^2\,g^{-1}}$ and $w=100\,\kms$ \citep{Jia+26}.

\begin{figure}[!htbp]
\centering
\includegraphics[width=0.83\linewidth]{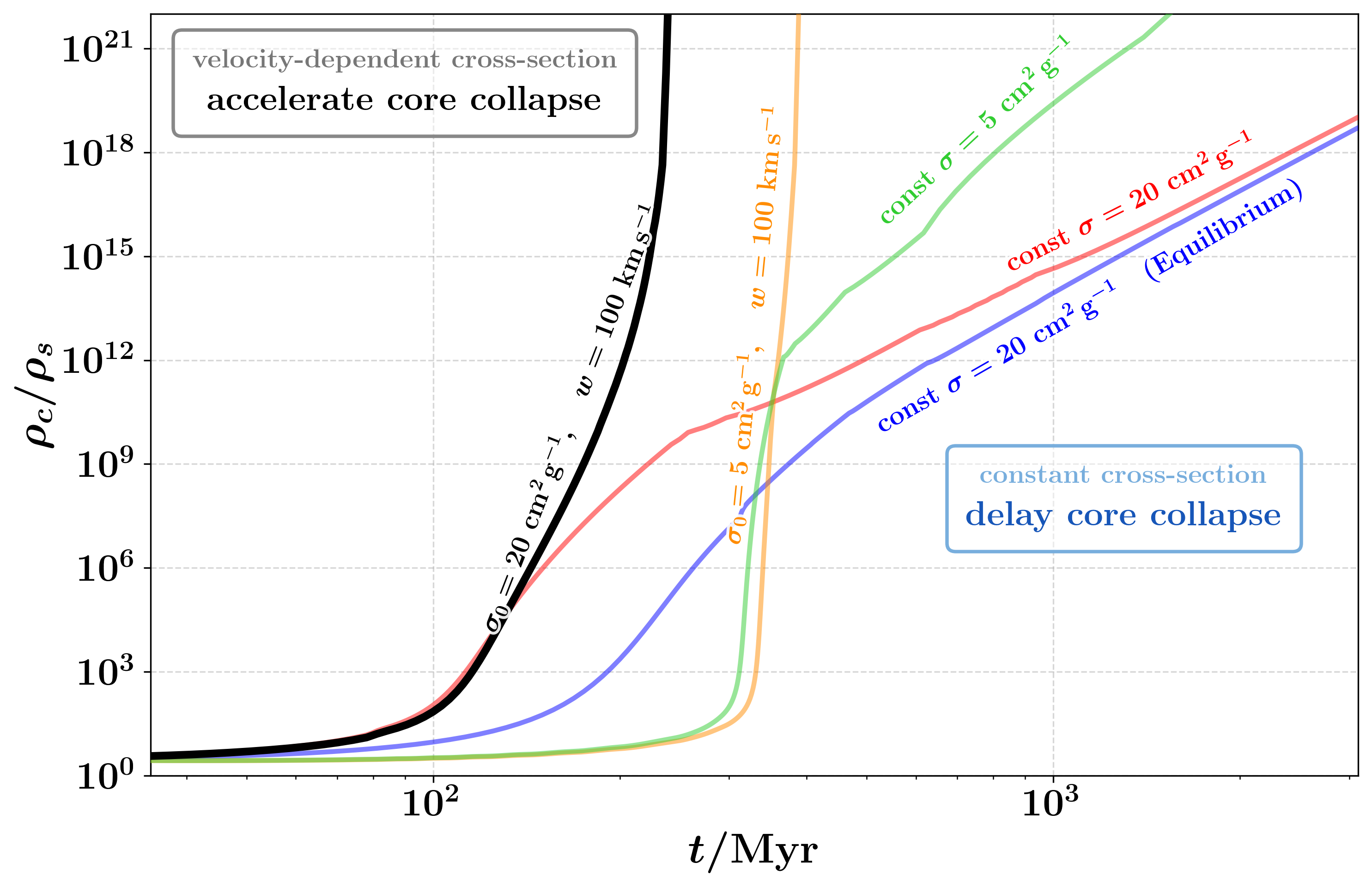}
\caption{
{\bf Impact of the self-interaction cross section and numerical method on gravothermal evolution} -- 
Evolution of the central density of an SIDM halo with $\Mv=10^8\Msun$ and $c=10$, initialized at $z=15$, for different self-interaction cross-section models. 
Models with velocity-dependent cross sections (black and orange) undergo accelerated late-stage gravothermal collapse, whereas models with constant cross sections (red and green) collapse more slowly at late times. 
The blue curve shows the conventional equilibrium treatment using the same cross-section parameters as the red curve.
}
\label{fig:cross_section}
\end{figure}

\Fig{cross_section} shows the central-density evolution of an SIDM halo for different cross-section models. The distinct collapse trajectories are governed by the transition between heat-conduction regimes. In the low-density, long-mean-free-path (LMFP) regime, where the dark-matter collisional mean free path greatly exceeds the characteristic halo scale height, the conductivity scales as $\kappa\propto\sigma$. As the core enters the high-density, short-mean-free-path (SMFP) regime at late times, however, the scaling changes to $\kappa\propto\sigma^{-1}$ \citep{Balberg+02a}.

For constant-cross-section models, a larger cross section (red) suppresses SMFP heat conduction more strongly than a smaller one (green), delaying late-time collapse and preventing timely BH-seed formation. Incorporating a velocity-dependent cross section naturally circumvents this bottleneck. As the core velocity dispersion increases, the effective cross section decreases as $v^{-4}$ in the SMFP regime. This reduction in $\sigma$ increases the conductivity and accelerates the final collapse phase. 

Among the velocity-dependent models, the one with larger $\sigma_0$ (black) collapses faster than that with smaller $\sigma_0$ (orange) because it maximizes the initial LMFP conductivity without suffering the late-time SMFP penalty. Under this optimal scenario, a seed BH forms within $\sim250\Myr$. Because astrophysical observations independently favor velocity-dependent cross sections [e.g., \citealp{Loeb+11,Kaplinghat+16}], this accelerated collapse pathway is observationally motivated.

\subsection*{Super-Bondi accretion}

After the BH seed forms, we follow its subsequent growth by
introducing an inner boundary representing the horizon. This boundary acts as a
strictly one-way membrane: mass and energy can cross it only inward, while the
horizon expands as it absorbs the inflowing material (\Fig{horizon}, upper
panel). 

The standard Bondi accretion rate is
\begin{equation}\label{eq:Eq6}
\dot{M}_{\mathrm{B}}
=4\pi\lambda_s G^2\Mbh^2\rho_\infty c_{\mathrm{s},\infty}^{-3},
\end{equation}
which scales as $\dot{M}_{\mathrm{B}}\propto\Mbh^2$ and becomes significant
only once the BH is sufficiently massive. During Stage 4, however, the
inflow is nearly in free fall, with its velocity approaching the speed of
light near the horizon, producing a mass flux far above the Bondi rate. Such
super-Bondi accretion does not violate the Eddington constraint, since SIDM
does not produce radiation and hence is not subject to the radiative Eddington limit
\citep{Feng+25}.

During early Stage 5, the accretion rate can exceed the Bondi limit by two orders of magnitude for an extended period. This super-Bondi phase is governed by the ratio of two timescales: the LMFP relaxation time, which characterizes SIDM particle collisions,
\begin{equation}
t_\mathrm{r}\sim
\left(\rho_\infty c_{\mathrm{s},\infty}\sigma_0\right)^{-1}
\approx7.5\Myr,
\end{equation}
and the Bondi accretion timescale,
\begin{equation}
t_\mathrm{B}\sim
\sqrt{\frac{R_\mathrm{s}^3}{G\Mbh}}.
\end{equation}
At the onset of Stage 5, $\Mbh=300\Msun$ and $t_\mathrm{B}\approx440\Myr\sim60t_\mathrm{r}$, so gravothermal collapse outpaces BH capture by a factor of $\sim60$, explaining the nearly two-order-of-magnitude accretion enhancement. As the BH grows to $\Mbh\approx10^6\Msun$, $t_\mathrm{B}$ falls to $\approx7.6\Myr\sim t_\mathrm{r}$; heat conduction no longer dominates, and the accretion rate approaches the Bondi limit (\Fig{accretion}). This transition is consistent with the dark Bondi accretion picture of \citep{Feng+25}, in which baryons accelerate early growth while dark-matter accretion sets the final BH mass. Adding baryons to our pure-SIDM model may thus shorten the growth timescale without substantially changing the final mass, which is primarily determined by the available dark-matter reservoir and therefore closely tied to the halo mass.

Toward the end of Stage 5, BH accretion reduces the halo mass and envelope density, lowering both the Bondi-predicted and actual accretion rates (\Fig{accretion}). After the BH consumes approximately $20\%$ of the halo mass, the envelope density drops to $\sim1\%$ of its initial value (\Fig{horizon}, upper panel, purple line), elongating the relaxation time to $t_\mathrm{r}\approx750\Myr$. SIDM collisions then become too infrequent on the evolutionary timescale to sustain the fluid approximation, placing subsequent evolution beyond the applicability of our fluid simulation. We therefore regard the final BH mass shown in \Fig{observation} as an upper limit.

\subsection*{Parameter space scanning}

We systematically scan $6.5<\log(\Mv/\mathrm{M}_{\odot})<9.5$ and $6<c<15$ to determine how the BH-seed formation and DM-accretion times depend on halo properties. The results are summarized in \Fig{scan}.

At the low-mass end, the BH-seed formation time (white dashed contours) increases
sharply because of the longer SIDM relaxation time. At higher masses, however,
the formation time becomes largely independent of mass and is governed
primarily by concentration. For the subsequent evolution, the total SMBH-formation timescale ---defined as the time at which the BH has consumed half of the isolated halo mass---depends nonmonotonically on halo mass. Although higher concentrations consistently accelerate accretion, a delay occurs near $\Mv\sim10^{8.5}\Msun$. This localized feature arises from an exceptionally low LMFP density produced by the velocity-dependent cross section\footnote{The LMFP density is established during Stage 2, as the core enters the SMFP regime. If the core velocity dispersion approaches the scale velocity $w$ at this transition, heat conduction becomes particularly efficient, producing a large density contrast between the inner SMFP core and the outer LMFP envelope. For $\Mv=10^{8.5}\Msun$, $v\sim100\,\kms$ at the transition.}. 

Our parameter scan shows that halos initialized at $z=15$ with $c\gtrsim10$ can generally produce SMBHs with $\Mbh\gtrsim10^6\Msun$ by $z\sim6$ (magenta contour). At the low-mass end, halos with $\Mv\lesssim10^7\Msun$ are disfavored by frequent halo mergers (grey dashed line, \Fig{merger}), whereas the halo abundance declines sharply at $\Mv\gtrsim10^9\Msun$ (purple dashed lines). The viable parameter space is therefore approximately $10^7\Msun\lesssim \Mv\lesssim10^9\Msun$ and $c\gtrsim10$.

We also test the sensitivity of these results to the cross section. For $\sigma_0=5\,\mathrm{cm^2\,g^{-1}}$ and $w=100\,\kms$, the gravothermal phases (Stages 1 and 2) last longer (\Fig{cross_section}). This longer evolution, however, allows the LMFP envelope to reach a higher density\footnote{Before the LMFP-to-SMFP transition, the LMFP density continues to increase. A smaller cross section delays this transition, after which the SMFP region continues to contract while the LMFP envelope remains nearly unchanged.}. Consequently, after the BH seed forms, its Stage 5 accretion rate is higher. The total evolution time (Stages 1--5) is close to that in the fiducial model ($\sigma_0=20\,\mathrm{cm^2\,g^{-1}}$ and $w=100\,\kms$). These results indicate that, over $\sigma_0=5$--$20\,\mathrm{cm^2\,g^{-1}}$, the predicted high-redshift SMBH number density is insensitive to the cross section.

\subsection*{Merger tree analysis}

\begin{figure}[!htbp]
\centering
\includegraphics[width=0.8\linewidth]{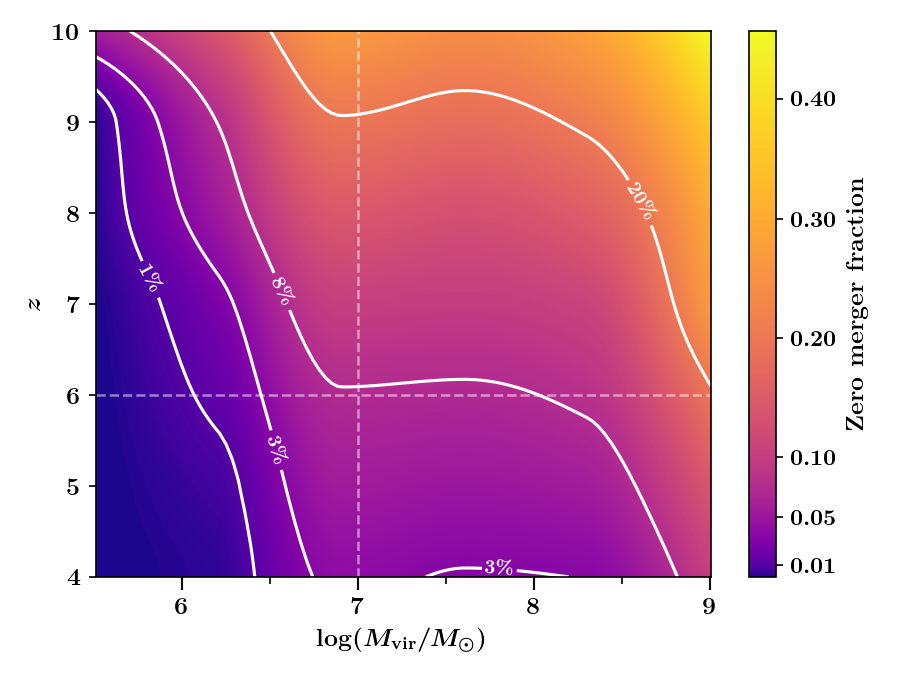}
\caption{
{\bf Exploring the halo mass and redshift regime in which gravothermal evolution is unlikely to be interrupted by major mergers} -- fraction of halos that experience no major merger (mass ratio $>1{:}3$) between $z=15$ and redshift $z$, as a function of halo mass, $\Mv$, at $z=15$. 
The fraction declines sharply below $\Mv\sim10^7\Msun$, indicating that low-mass halos are unlikely to complete BH seeding before being disrupted by major mergers. 
Among the halos with $10^7\Msun\lesssim\Mv\lesssim10^8\Msun$, approximately $8\%$ remain free of major mergers until $z=6$, allowing them to complete gravothermal evolution and form a supermassive BH by the LRD epoch.
}
\label{fig:merger}
\end{figure}

Only a small fraction of viable halos ($\Mv\gtrsim10^7\Msun$ and $c\gtrsim10$
at $z=15$) can form high-redshift SMBHs because strong external energy
injection---primarily from major mergers---can interrupt their formation. To
quantify the fraction of viable halos affected by major mergers during SMBH
formation, we construct cosmological merger trees using the \textit{SatGen}
algorithm \citep{Jiang+20,Green+21a,Green+21b}. 
We adopt a final redshift $z=4$
and generate $10000$ merger trees for halos with final masses in the range
$10^8\Msun<\Mv<10^{11}\Msun$. We then analyze the number of major mergers
between redshifts $z$ and $15$ as a function of halo virial mass at $z=15$.
Specifically, we determine the fraction of halos that undergo no mergers with a
mass ratio larger than 1:3 over this redshift interval. The results are shown
in \Fig{merger}.

As shown by the contours, only $\gtrsim8\%$ of halos have undergone no major merger by $z=6$ and therefore successfully form a high-redshift SMBH. Within the viable mass range $\Mv\gtrsim10^7\Msun$, this fraction is insensitive to halo mass. Below $\Mv\sim10^7\Msun$, however, this fraction decreases sharply. Thus, even given sufficient time, low-mass halos cannot form SMBHs because frequent major mergers disrupt their formation. Moreover, by $z=4$, most halos have undergone at least one major merger. Therefore, the SIDM scenario naturally suppresses low-redshift SMBH seeding, consistent with the observed LRD redshift range $4\lesssim z\lesssim9$.

\FloatBarrier
\backmatter

\bmhead{Acknowledgements}

We thank Wei-Xiang Feng, Hai-bo Yu, Frank van den Bosch, and Zi-Hao Wang for helpful discussions. We thank Luis Ho for his encouragement and support of this work. This work was supported by the National Natural Science Foundation of China (Grant No. 12473037). FJ acknowledges support by the National Natural Science Foundation of China (NSFC, 12473007) and China Manned Space Program with grant no. CMS-CSST-2025-A03. RL acknowledges the support of National Nature Science Foundation of China (No 11988101), the science research grants from the China Manned Space Program (No. CMS-CSST-2025-A0), CAS Project for Young Scientists in Basic Research (No. YSBR-062).

\bmhead{Contributions}

H.P.G. developed and extended the Misner-Sharp framework, performed the calculations presented in this work, generated the figures, and prepared the first draft. 
F.J. co-conceived the project, advised on the cosmic-structure-formation aspects of this work, and developed the manuscript's narrative and logical structure. 
X.C. proposed and conceived this project, advised on the development of the extended Misner-Sharp framework, and co-developed the narrative and logical structure of the manuscript. 
R.L. provided key input on the interpretations of the results and the figure comparing the model with observations. 
Z.X.J. constructed and analyzed halo merger trees used in this work.
All authors contributed to the interpretation of the results, provided feedback on the analysis and manuscript, and approved the final version.

\bmhead{Code vailability}
The code used to model SIDM halo evolution in the Misner-Sharp framework is available in Ref.~\cite{Gu+26b} and can be accessed at \href{https://github.com/Hua-Peng-G/SIDM}{https://github.com/Hua-Peng-G/SIDM}.

\bibliography{refs}

@PREAMBLE{
 "\providecommand{\noopsort}[1]{}" 
 # "\providecommand{\singleletter}[1]{#1}%" 
}

@article{Adhikari+22,
  author = "Adhikari, Susmita and others",
  title = "{Astrophysical Tests of Dark Matter Self-Interactions}",
  eprint = "2207.10638",
  archiveprefix = "arXiv",
  primaryclass = "astro-ph.CO",
  month = {12},
  year = {2025},
  doi = {10.1103/m2vm-59y3},
  journal = {Rev. Mod. Phys.},
  volume = {97},
  number = {4},
  pages = {045004},
  url = {https://doi.org/10.1103/m2vm-59y3}
}

@article{Balberg+02a,
  author = "Balberg, Shmuel and Shapiro, Stuart L. and Inagaki, Shogo",
  title = "{Selfinteracting dark matter halos and the gravothermal catastrophe}",
  eprint = "astro-ph/0110561",
  archiveprefix = "arXiv",
  doi = "10.1086/339038",
  journal = "Astrophys. J.",
  volume = "568",
  pages = "475--487",
  year = "2002",
  url = {https://doi.org/10.1086/339038}
}

@article{Barack+19,
  author = "Barack, Leor and others",
  title = "{Black holes, gravitational waves and fundamental physics: a roadmap}",
  eprint = "1806.05195",
  archiveprefix = "arXiv",
  primaryclass = "gr-qc",
  doi = "10.1088/1361-6382/ab0587",
  journal = "Class. Quant. Grav.",
  volume = "36",
  number = "14",
  pages = "143001",
  year = "2019",
  url = {https://doi.org/10.1088/1361-6382/ab0587}
}

@article{Barkana+01,
  title = {In the beginning: the first sources of light and the reionization of the universe},
  journal = {Physics Reports},
  volume = {349},
  number = {2},
  pages = {125-238},
  year = {2001},
  issn = {0370-1573},
  doi = {10.1016/S0370-1573(01)00019-9},
  url = {https://doi.org/10.1016/S0370-1573(01)00019-9},
  author = {Rennan Barkana and Abraham Loeb}
}

@article{Begelman+06,
  author = "Begelman, Mitchell C. and Volonteri, Marta and Rees, Martin J.",
  title = "{Formation of supermassive black holes by direct collapse in pregalactic halos}",
  eprint = "astro-ph/0602363",
  archiveprefix = "arXiv",
  doi = "10.1111/j.1365-2966.2006.10467.x",
  journal = "Mon. Not. Roy. Astron. Soc.",
  volume = "370",
  pages = "289--298",
  year = "2006",
  url = {https://doi.org/10.1111/j.1365-2966.2006.10467.x}
}

@article{Bromm+03,
  author = "Bromm, Volker and Loeb, Abraham",
  title = "{Formation of the first supermassive black holes}",
  eprint = "astro-ph/0212400",
  archiveprefix = "arXiv",
  doi = "10.1086/377529",
  journal = "Astrophys. J.",
  volume = "596",
  pages = "34--46",
  year = "2003",
  url = {https://doi.org/10.1086/377529}
}

@ARTICLE{Cao+26,
  author = {{Cao}, Xiaoyue and others},
  title = "{Probing dark matter substructures with free-form modelling: a case study of the 'Jackpot' strong lens}",
  journal = {\mnras},
  year = 2026,
  month = feb,
  volume = {545},
  number = {4},
  eid = {staf2179},
  pages = {staf2179},
  doi = {10.1093/mnras/staf2179},
  archiveprefix = {arXiv},
  eprint = {2504.19177},
  primaryclass = {astro-ph.CO},
  adsurl = {https://ui.adsabs.harvard.edu/abs/2026MNRAS.545f2179C},
  url = {https://doi.org/10.1093/mnras/staf2179}
}

@article{Chantavat+23,
  author = "Chantavat, Teeraparb and Chongchitnan, Siri and Silk, Joseph",
  title = "{The most massive Population III stars}",
  eprint = "2302.09763",
  archiveprefix = "arXiv",
  primaryclass = "astro-ph.SR",
  doi = "10.1093/mnras/stad1196",
  journal = "Mon. Not. Roy. Astron. Soc.",
  volume = "522",
  number = "3",
  pages = "3256--3262",
  year = "2023",
  url = {https://doi.org/10.1093/mnras/stad1196}
}

@article{Chen+25,
  author = "Chen, Chang-Hao and Ho, Luis C. and Li, Ruancun and Zhuang, Ming-Yang",
  title = "{The Host Galaxy (If Any) of the Little Red Dots}",
  doi = "10.3847/1538-4357/ada93a",
  journal = "Astrophys. J.",
  volume = "983",
  number = "1",
  pages = "60",
  year = "2025",
  url = {https://doi.org/10.3847/1538-4357/ada93a}
}

@article{Choquette+19,
  author = "Choquette, Jeremie and Cline, James M. and Cornell, Jonathan M.",
  title = "{Early formation of supermassive black holes via dark matter self-interactions}",
  eprint = "1812.05088",
  archiveprefix = "arXiv",
  primaryclass = "astro-ph.CO",
  doi = "10.1088/1475-7516/2019/07/036",
  journal = "JCAP",
  volume = "07",
  pages = "036",
  year = "2019",
  url = {https://doi.org/10.1088/1475-7516/2019/07/036}
}

@ARTICLE{Dekel+23,
  author = {{Dekel}, Avishai and {Sarkar}, Kartick C. and {Birnboim}, Yuval and {Mandelker}, Nir and {Li}, Zhaozhou},
  title = "{Efficient formation of massive galaxies at cosmic dawn by feedback-free starbursts}",
  journal = {\mnras},
  year = 2023,
  month = aug,
  volume = {523},
  number = {3},
  pages = {3201-3218},
  doi = {10.1093/mnras/stad1557},
  archiveprefix = {arXiv},
  eprint = {2303.04827},
  primaryclass = {astro-ph.GA},
  adsurl = {https://ui.adsabs.harvard.edu/abs/2023MNRAS.523.3201D},
  url = {https://doi.org/10.1093/mnras/stad1557}
}

@article{Eckart+40,
  author = "Eckart, Carl",
  title = "{The Thermodynamics of irreversible processes. 3.. Relativistic theory of the simple fluid}",
  doi = "10.1103/PhysRev.58.919",
  journal = "Phys. Rev.",
  volume = "58",
  pages = "919--924",
  year = "1940",
  url = {https://doi.org/10.1103/PhysRev.58.919}
}

@ARTICLE{Feng+09,
  author = {{Feng}, Jonathan L. and {Kaplinghat}, Manoj and {Tu}, Huitzu and {Yu}, Hai-Bo},
  title = "{Hidden charged dark matter}",
  journal = {\jcap},
  year = 2009,
  month = jul,
  volume = {2009},
  number = {7},
  eid = {004},
  pages = {004},
  doi = {10.1088/1475-7516/2009/07/004},
  archiveprefix = {arXiv},
  eprint = {0905.3039},
  primaryclass = {hep-ph},
  adsurl = {https://ui.adsabs.harvard.edu/abs/2009JCAP...07..004F},
  url = {https://doi.org/10.1088/1475-7516/2009/07/004}
}

@article{Feng+21,
  author = "Feng, Wei-Xiang and Yu, Hai-Bo and Zhong, Yi-Ming",
  title = "{Seeding Supermassive Black Holes with Self-interacting Dark Matter: A Unified Scenario with Baryons}",
  eprint = "2010.15132",
  archiveprefix = "arXiv",
  primaryclass = "astro-ph.CO",
  doi = "10.3847/2041-8213/ac04b0",
  journal = "Astrophys. J. Lett.",
  volume = "914",
  number = "2",
  pages = "L26",
  year = "2021",
  url = {https://doi.org/10.3847/2041-8213/ac04b0}
}

@article{Feng+22,
  author = "Feng, Wei-Xiang and Yu, Hai-Bo and Zhong, Yi-Ming",
  title = "{Dynamical instability of collapsed dark matter halos}",
  eprint = "2108.11967",
  archiveprefix = "arXiv",
  primaryclass = "astro-ph.CO",
  doi = "10.1088/1475-7516/2022/05/036",
  journal = "JCAP",
  volume = "05",
  number = "05",
  pages = "036",
  year = "2022",
  url = {https://doi.org/10.1088/1475-7516/2022/05/036}
}

@article{Feng+25,
  author = "Feng, Wei-Xiang and Yu, Hai-Bo and Zhong, Yi-Ming",
  title = "{Dark Bondi Accretion Aided by Baryons and the Origin of JWST Little Red Dots}",
  eprint = "2506.17641",
  archiveprefix = "arXiv",
  primaryclass = "astro-ph.GA",
  month = "6",
  year = "2025",
  url = {https://arxiv.org/abs/2506.17641}
}

@article{Fischer+24,
  author = "Fischer, Moritz S. and Dolag, Klaus and Yu, Hai-Bo",
  title = "{Numerical challenges for energy conservation in N-body simulations of collapsing self-interacting dark matter halos}",
  eprint = "2403.00739",
  archiveprefix = "arXiv",
  primaryclass = "astro-ph.CO",
  doi = "10.1051/0004-6361/202449849",
  journal = "Astron. Astrophys.",
  volume = "689",
  pages = "A300",
  year = "2024",
  url = {https://doi.org/10.1051/0004-6361/202449849}
}

@article{Gad-Nasr+24,
  author = "Gad-Nasr, Sophia and Boddy, Kimberly K. and Kaplinghat, Manoj and Outmezguine, Nadav Joseph and Sagunski, Laura",
  title = "{On the late-time evolution of velocity-dependent self-interacting dark matter halos}",
  eprint = "2312.09296",
  archiveprefix = "arXiv",
  primaryclass = "astro-ph.GA",
  doi = "10.1088/1475-7516/2024/05/131",
  journal = "JCAP",
  volume = "05",
  pages = "131",
  year = "2024",
  url = {https://doi.org/10.1088/1475-7516/2024/05/131}
}

@article{Gilman+21b,
  author = "Gilman, Daniel and Bovy, Jo and Treu, Tommaso and Nierenberg, Anna and Birrer, Simon and Benson, Andrew and Sameie, Omid",
  title = "{Strong lensing signatures of self-interacting dark matter in low-mass haloes}",
  eprint = "2105.05259",
  archiveprefix = "arXiv",
  primaryclass = "astro-ph.CO",
  doi = "10.1093/mnras/stab2335",
  journal = "Mon. Not. Roy. Astron. Soc.",
  volume = "507",
  number = "2",
  pages = "2432--2447",
  year = "2021",
  url = {https://doi.org/10.1093/mnras/stab2335}
}

@ARTICLE{Green+21a,
  author = {{Green}, Sheridan B. and {van den Bosch}, Frank C. and {Jiang}, Fangzhou},
  title = "{The tidal evolution of dark matter substructure - II. The impact of artificial disruption on subhalo mass functions and radial profiles}",
  journal = {\mnras},
  year = 2021,
  month = may,
  volume = {503},
  number = {3},
  pages = {4075-4091},
  doi = {10.1093/mnras/stab696},
  archiveprefix = {arXiv},
  eprint = {2103.01227},
  primaryclass = {astro-ph.GA},
  adsurl = {https://ui.adsabs.harvard.edu/abs/2021MNRAS.503.4075G},
  url = {https://doi.org/10.1093/mnras/stab696}
}

@ARTICLE{Green+21b,
  author = {{Green}, Sheridan B. and {van den Bosch}, Frank C. and {Jiang}, Fangzhou},
  title = "{SatGen - II. Assessing the impact of a disc potential on subhalo populations}",
  journal = {\mnras},
  year = 2022,
  month = jan,
  volume = {509},
  number = {2},
  pages = {2624-2636},
  doi = {10.1093/mnras/stab3130},
  archiveprefix = {arXiv},
  eprint = {2110.13044},
  primaryclass = {astro-ph.GA},
  adsurl = {https://ui.adsabs.harvard.edu/abs/2022MNRAS.509.2624G},
  url = {https://doi.org/10.1093/mnras/stab3130}
}

@article{Greene+24,
  author = "Greene, Jenny E. and others",
  title = "{UNCOVER Spectroscopy Confirms the Surprising Ubiquity of Active Galactic Nuclei in Red Sources at z \ensuremath{>} 5}",
  doi = "10.3847/1538-4357/ad1e5f",
  journal = "Astrophys. J.",
  volume = "964",
  number = "1",
  pages = "39",
  year = "2024",
  url = {https://doi.org/10.3847/1538-4357/ad1e5f}
}

@ARTICLE{Greene+26,
  author = {{Greene}, Jenny E. and {Setton}, David J. and {Furtak}, Lukas J. and {Naidu}, Rohan P. and {Volonteri}, Marta and {Dayal}, Pratika and {Labbe}, Ivo and {van Dokkum}, Pieter and {Bezanson}, Rachel and {Brammer}, Gabriel and {Cutler}, Sam E. and {Glazebrook}, Karl and {de Graaff}, Anna and {Hirschmann}, Michaela and {Hviding}, Raphael E. and {Kokorev}, Vasily and {Leja}, Joel and {Liu}, Hanpu and {Ma}, Yilun and {Matthee}, Jorryt and {Nanayakkara}, Themiya and {Oesch}, Pascal A. and {Pan}, Richard and {Price}, Sedona H. and {Spilker}, Justin S. and {Wang}, Bingjie and {Weaver}, John R. and {Whitaker}, Katherine E. and {Williams}, Christina C. and {Zitrin}, Adi},
  title = "{What You See Is What You Get: Empirically Measured Bolometric Luminosities of Little Red Dots}",
  journal = {\apj},
  year = 2026,
  month = jan,
  volume = {996},
  number = {2},
  eid = {129},
  pages = {129},
  doi = {10.3847/1538-4357/ae1836},
  archiveprefix = {arXiv},
  eprint = {2509.05434},
  primaryclass = {astro-ph.GA},
  adsurl = {https://ui.adsabs.harvard.edu/abs/2026ApJ...996..129G},
  url = {https://doi.org/10.3847/1538-4357/ae1836}
}

@article{Gu+26a,
  author = "Gu, Hua-Peng and Jiang, Fangzhou and Chen, Xian and Li, Ran",
  title = "{Nonequilibrium relativistic core collapse of self-interacting dark matter halos: Limits on the seed black hole mass}",
  eprint = "2601.17117",
  archiveprefix = "arXiv",
  primaryclass = "astro-ph.CO",
  doi = "10.1103/6sl3-kjzf",
  journal = "Phys. Rev. D",
  volume = "113",
  number = "10",
  pages = "103038",
  year = "2026",
  url = {https://doi.org/10.1103/6sl3-kjzf}
}

@misc{Gu+26b,
  author = {Gu, Huapeng. and Jiang, Fangzhou. and Chen, Xian.},
  title = {{SIDM Halo Collapse Simulation in GR + Hydrodynamical}},
  year = {2026},
  publisher = {GitHub},
  journal = {GitHub repository},
  url = {https://github.com/Hua-Peng-G/SIDM}
}

@book{Hawking+23,
  author = "Hawking, Stephen W. and Ellis, George F. R.",
  title = "{The Large Scale Structure of Space-Time}",
  doi = "10.1017/9781009253161",
  isbn = "978-1-009-25316-1, 978-1-009-25315-4, 978-0-521-20016-5, 978-0-521-09906-6, 978-0-511-82630-6, 978-0-521-09906-6",
  publisher = "Cambridge University Press",
  series = "Cambridge Monographs on Mathematical Physics",
  month = "2",
  year = "2023",
  url = {https://doi.org/10.1017/9781009253161}
}

@article{Hernandez+66,
  author = "Hernandez, Walter C. and Misner, Charles W.",
  title = "{Observer Time as a Coordinate in Relativistic Spherical Hydrodynamics}",
  doi = "10.1086/148525",
  journal = "Astrophys. J.",
  volume = "143",
  pages = "452",
  year = "1966",
  url = {https://doi.org/10.1086/148525}
}

@ARTICLE{Hu+06a,
  author = {{Hu}, Jian and {Shen}, Yue and {Lou}, Yu-Qing and {Zhang}, Shuangnan},
  title = "{Forming supermassive black holes by accreting dark and baryon matter}",
  journal = {\mnras},
  year = 2006,
  month = jan,
  volume = {365},
  number = {1},
  pages = {345-351},
  doi = {10.1111/j.1365-2966.2005.09712.x},
  archiveprefix = {arXiv},
  eprint = {astro-ph/0510222},
  primaryclass = {astro-ph},
  adsurl = {https://ui.adsabs.harvard.edu/abs/2006MNRAS.365..345H},
  url = {https://doi.org/10.1111/j.1365-2966.2005.09712.x}
}

@article{Inayoshi+20,
  author = "Inayoshi, Kohei and Visbal, Eli and Haiman, Zolt\'an",
  title = "{The Assembly of the First Massive Black Holes}",
  eprint = "1911.05791",
  archiveprefix = "arXiv",
  primaryclass = "astro-ph.GA",
  doi = "10.1146/annurev-astro-120419-014455",
  journal = "Ann. Rev. Astron. Astrophys.",
  volume = "58",
  pages = "27--97",
  year = "2020",
  url = {https://doi.org/10.1146/annurev-astro-120419-014455}
}

@article{Ivey+26,
  author = "Ivey, Lucy R. and others",
  title = "{The Cliff: A Metal-Poor Little Red Dot Hosting an Overmassive Black Hole at $z=3.55$}",
  eprint = "2604.09177",
  archiveprefix = "arXiv",
  primaryclass = "astro-ph.GA",
  journal = {Mon. Not. Roy. Astron. Soc.},
  year = {2026},
  doi = {10.1093/mnras/stag1220},
  volume = {550},
  number = {4},
  pages = {stag1220},
  url = {https://doi.org/10.1093/mnras/stag1220}
}

@ARTICLE{Jia+26,
  author = {{Jia}, Zixiang and {Jiang}, Fangzhou and {Li}, Shubo and {Li}, Ran and {Wang}, Jing and {Zhu}, Ling},
  title = "{An Enhanced Isothermal Jeans Approach to Constraining Dark Matter Self-Interactions from Galactic Kinematics}",
  journal = {\mnras},
  year = 2026,
  month = may,
  doi = {10.1093/mnras/stag969},
  archiveprefix = {arXiv},
  eprint = {2601.17118},
  primaryclass = {astro-ph.GA},
  adsurl = {https://ui.adsabs.harvard.edu/abs/2026MNRAS.tmp..911J},
  url = {https://doi.org/10.1093/mnras/stag969}
}

@ARTICLE{Jiang+20,
  author = {{Jiang}, Fangzhou and {Dekel}, Avishai and {Freundlich}, Jonathan and {van den Bosch}, Frank C. and {Green}, Sheridan B. and {Hopkins}, Philip F. and {Benson}, Andrew and {Du}, Xiaolong},
  title = "{SatGen: a semi-analytical satellite galaxy generator - I. The model and its application to Local-Group satellite statistics}",
  journal = {\mnras},
  year = 2021,
  month = mar,
  volume = {502},
  number = {1},
  pages = {621-641},
  doi = {10.1093/mnras/staa4034},
  archiveprefix = {arXiv},
  eprint = {2005.05974},
  primaryclass = {astro-ph.GA},
  adsurl = {https://ui.adsabs.harvard.edu/abs/2021MNRAS.502..621J},
  url = {https://doi.org/10.1093/mnras/staa4034}
}

@article{Jiang+23,
  author = "Jiang, Fangzhou and Benson, Andrew and Hopkins, Philip F. and Slone, Oren and Lisanti, Mariangela and Kaplinghat, Manoj and Peter, Annika H. G. and Zeng, Zhichao Carton and Du, Xiaolong and Yang, Shengqi and Shen, Xuejian",
  title = "{A semi-analytic study of self-interacting dark-matter haloes with baryons}",
  eprint = "2206.12425",
  archiveprefix = "arXiv",
  primaryclass = "astro-ph.CO",
  doi = "10.1093/mnras/stad705",
  journal = "Mon. Not. Roy. Astron. Soc.",
  volume = "521",
  number = "3",
  pages = "4630--4644",
  year = "2023",
  url = {https://doi.org/10.1093/mnras/stad705}
}

@article{Jiang+25,
  author = "Jiang, Fangzhou and Jia, Zixiang and Zheng, Haonan and Ho, Luis C. and Inayoshi, Kohei and Shen, Xuejian and Vogelsberger, Mark and Feng, Wei-Xiang",
  title = "{Formation of the Little Red Dots from the Core-collapse of Self-interacting Dark Matter Halos}",
  eprint = "2503.23710",
  archiveprefix = "arXiv",
  primaryclass = "astro-ph.GA",
  month = {1},
  year = {2026},
  doi = {10.3847/2041-8213/ae247a},
  journal = {Astrophys. J. Lett.},
  volume = {996},
  number = {1},
  pages = {L19},
  url = {https://doi.org/10.3847/2041-8213/ae247a}
}

@ARTICLE{Jones+25,
  author = {{Jones}, Brenda L. and others},
  title = "{The $M_{\rm BH}-M_{*}$ Relationship at $3<z<7$: Big Black Holes in Little Red Dots}",
  journal = {arXiv e-prints},
  year = 2025,
  month = oct,
  eid = {arXiv:2510.07376},
  pages = {arXiv:2510.07376},
  doi = {10.48550/arXiv.2510.07376},
  archiveprefix = {arXiv},
  eprint = {2510.07376},
  primaryclass = {astro-ph.GA},
  adsurl = {https://ui.adsabs.harvard.edu/abs/2025arXiv251007376J},
  url = {https://arxiv.org/abs/2510.07376}
}

@ARTICLE{Juodz+26,
  author = {{Juod{\v{z}}balis}, Ignas and others},
  title = "{A direct black-hole mass measurement in a little red dot at high redshift}",
  journal = {\nat},
  year = 2026,
  month = may,
  volume = {653},
  number = {8116},
  pages = {1017-1021},
  doi = {10.1038/s41586-026-10579-4},
  archiveprefix = {arXiv},
  eprint = {2508.21748},
  primaryclass = {astro-ph.GA},
  adsurl = {https://ui.adsabs.harvard.edu/abs/2026Natur.653.1017J},
  url = {https://doi.org/10.1038/s41586-026-10579-4}
}

@ARTICLE{Kaplinghat+16,
  author = {{Kaplinghat}, Manoj and {Tulin}, Sean and {Yu}, Hai-Bo},
  title = "{Dark Matter Halos as Particle Colliders: Unified Solution to Small-Scale Structure Puzzles from Dwarfs to Clusters}",
  journal = {\prl},
  year = 2016,
  month = jan,
  volume = {116},
  number = {4},
  eid = {041302},
  pages = {041302},
  doi = {10.1103/PhysRevLett.116.041302},
  archiveprefix = {arXiv},
  eprint = {1508.03339},
  primaryclass = {astro-ph.CO},
  adsurl = {https://ui.adsabs.harvard.edu/abs/2016PhRvL.116d1302K},
  url = {https://doi.org/10.1103/PhysRevLett.116.041302}
}

@ARTICLE{Kocevski+24,
  author = {{Kocevski}, Dale D. and others},
  title = "{The Rise of Faint, Red AGN at $z>4$: A Sample of Little Red Dots in the JWST Extragalactic Legacy Fields}",
  journal = {Astrophys. J.},
  year = {2025},
  month = {6},
  eid = {126},
  pages = {126},
  doi = {10.3847/1538-4357/adbc7d},
  archiveprefix = {arXiv},
  eprint = {2404.03576},
  primaryclass = {astro-ph.GA},
  adsurl = {https://ui.adsabs.harvard.edu/abs/2024arXiv240403576K},
  volume = {986},
  number = {2},
  url = {https://doi.org/10.3847/1538-4357/adbc7d}
}

@article{Koda+11,
  author = "Koda, Jun and Shapiro, Paul R.",
  title = "{Gravothermal collapse of isolated self-interacting dark matter haloes: N-body simulation versus the fluid model}",
  eprint = "1101.3097",
  archiveprefix = "arXiv",
  primaryclass = "astro-ph.CO",
  reportnumber = "TCC-001-11",
  doi = "10.1111/j.1365-2966.2011.18684.x",
  journal = "Mon. Not. Roy. Astron. Soc.",
  volume = "415",
  pages = "1125",
  year = "2011",
  url = {https://doi.org/10.1111/j.1365-2966.2011.18684.x}
}

@ARTICLE{Kokorev+24,
  author = {{Kokorev}, Vasily and others},
  title = "{A Census of Photometrically Selected Little Red Dots at 4 < z < 9 in JWST Blank Fields}",
  journal = {\apj},
  year = 2024,
  month = jun,
  volume = {968},
  number = {1},
  eid = {38},
  pages = {38},
  doi = {10.3847/1538-4357/ad4265},
  archiveprefix = {arXiv},
  eprint = {2401.09981},
  primaryclass = {astro-ph.GA},
  adsurl = {https://ui.adsabs.harvard.edu/abs/2024ApJ...968...38K},
  url = {https://doi.org/10.3847/1538-4357/ad4265}
}

@article{Kritos+24,
  author = "Kritos, Konstantinos and Berti, Emanuele and Silk, Joseph",
  title = "{Supermassive black holes from runaway mergers and accretion in nuclear star clusters}",
  eprint = "2404.11676",
  archiveprefix = "arXiv",
  primaryclass = "astro-ph.HE",
  doi = "10.1093/mnras/stae1145",
  journal = "Mon. Not. Roy. Astron. Soc.",
  volume = "531",
  number = "1",
  pages = "133--136",
  year = "2024",
  url = {https://doi.org/10.1093/mnras/stae1145}
}

@article{Latif+13,
  author = "Latif, M. A. and Schleicher, D. R. G. and Schmidt, W. and Niemeyer, J.",
  title = "{Black hole formation in the early universe}",
  eprint = "1304.0962",
  archiveprefix = "arXiv",
  primaryclass = "astro-ph.CO",
  doi = "10.1093/mnras/stt834",
  journal = "Mon. Not. Roy. Astron. Soc.",
  volume = "433",
  pages = "1607",
  year = "2013",
  url = {https://doi.org/10.1093/mnras/stt834}
}

@article{Latif+16,
  author = "Latif, Muhammad A. and Ferrara, Andrea",
  title = "{Formation of supermassive black hole seeds}",
  eprint = "1605.07391",
  archiveprefix = "arXiv",
  primaryclass = "astro-ph.GA",
  doi = "10.1017/pasa.2016.41",
  journal = "Publ. Astron. Soc. Austral.",
  volume = "33",
  pages = "e051",
  year = "2016",
  url = {https://doi.org/10.1017/pasa.2016.41}
}

@article{Li+25,
  author = "Li, Shubo and others",
  title = "{The {\textquotedblleft}Little Dark Dot{\textquotedblright}: Evidence for Self-interacting Dark Matter in the Strong Lens SDSS J0946+1006?}",
  eprint = "2504.11800",
  archiveprefix = "arXiv",
  primaryclass = "astro-ph.GA",
  doi = "10.3847/1538-4357/ae1462",
  journal = "Astrophys. J.",
  volume = "994",
  number = "2",
  pages = "201",
  year = "2025",
  url = {https://doi.org/10.3847/1538-4357/ae1462}
}

@ARTICLE{Loeb+11,
  author = {{Loeb}, Abraham and {Weiner}, Neal},
  title = "{Cores in Dwarf Galaxies from Dark Matter with a Yukawa Potential}",
  journal = {\prl},
  year = 2011,
  month = apr,
  volume = {106},
  number = {17},
  eid = {171302},
  pages = {171302},
  doi = {10.1103/PhysRevLett.106.171302},
  archiveprefix = {arXiv},
  eprint = {1011.6374},
  primaryclass = {astro-ph.CO},
  adsurl = {https://ui.adsabs.harvard.edu/abs/2011PhRvL.106q1302L},
  url = {https://doi.org/10.1103/PhysRevLett.106.171302}
}

@article{Loiacono+26,
  author = "Loiacono, Federica and others",
  title = "{No Evolution in the Number Density of Little Red Dots from Cosmic Dawn to Cosmic Noon}",
  eprint = "2606.30253",
  archiveprefix = "arXiv",
  primaryclass = "astro-ph.GA",
  journal = "arXiv e-prints",
  year = "2026",
  url = {https://arxiv.org/abs/2606.30253}
}

@article{Lora-Clavijo+14,
  author = "Lora-Clavijo, F. D. and Gracia-Linares, M. and Guzman, F. S.",
  title = "{Horizon growth of supermassive black hole seeds fed with collisional dark matter}",
  eprint = "1406.7233",
  archiveprefix = "arXiv",
  primaryclass = "astro-ph.GA",
  doi = "10.1093/mnras/stu1289",
  journal = "Mon. Not. Roy. Astron. Soc.",
  volume = "443",
  number = "3",
  pages = "2242--2251",
  year = "2014",
  url = {https://doi.org/10.1093/mnras/stu1289}
}

@ARTICLE{Maiolino+24,
  author = {{Maiolino}, Roberto and others},
  title = "{JADES: The diverse population of infant black holes at 4 < z < 11: Merging, tiny, poor, but mighty}",
  journal = {\aap},
  year = 2024,
  month = nov,
  volume = {691},
  eid = {A145},
  pages = {A145},
  doi = {10.1051/0004-6361/202347640},
  archiveprefix = {arXiv},
  eprint = {2308.01230},
  primaryclass = {astro-ph.GA},
  adsurl = {https://ui.adsabs.harvard.edu/abs/2024A&A...691A.145M},
  url = {https://doi.org/10.1051/0004-6361/202347640}
}

@article{Maiolino+26,
  author = "Maiolino, Roberto and others",
  title = "{A Black Hole in a Near-Pristine Galaxy 700 Million Years after the Big Bang}",
  eprint = "2505.22567",
  archiveprefix = "arXiv",
  primaryclass = "astro-ph.GA",
  doi = "10.1093/mnras/staf2109",
  journal = "Mon. Not. Roy. Astron. Soc.",
  volume = "548",
  number = "1",
  pages = "staf2109",
  year = "2026",
  url = {https://doi.org/10.1093/mnras/staf2109}
}

@article{Matthee+24,
  author = "Matthee, Jorryt and others",
  title = "{Little Red Dots: An Abundant Population of Faint Active Galactic Nuclei at z \ensuremath{\sim} 5 Revealed by the EIGER and FRESCO JWST Surveys}",
  eprint = "2306.05448",
  archiveprefix = "arXiv",
  primaryclass = "astro-ph.GA",
  doi = "10.3847/1538-4357/ad2345",
  journal = "Astrophys. J.",
  volume = "963",
  number = "2",
  pages = "129",
  year = "2024",
  url = {https://doi.org/10.3847/1538-4357/ad2345}
}

@online{Meng+26,
  title = {Spherically {{Symmetric Fluid Simulations}} of {{Black Hole Accretion}} in {{Self-Interacting Dark Matter Halos}}},
  author = {Meng, Zhe and Chen, Tan and Zhu, Bocheng and Zhou, Fan and Hu, Bin and Gao, Liang and Cai, Rong-Gen},
  date = {2026-07-02},
  eprint = {2607.02151},
  eprinttype = {arXiv},
  eprintclass = {astro-ph.CO},
  doi = {10.48550/arXiv.2607.02151},
  url = {https://arxiv.org/abs/2607.02151},
  urldate = {2026-07-05},
  pubstate = {prepublished}
}

@ARTICLE{Minor+21,
  author = {{Minor}, Quinn and {Gad-Nasr}, Sophia and {Kaplinghat}, Manoj and {Vegetti}, Simona},
  title = "{An unexpected high concentration for the dark substructure in the gravitational lens SDSSJ0946+1006}",
  journal = {\mnras},
  year = 2021,
  month = oct,
  volume = {507},
  number = {2},
  pages = {1662-1683},
  doi = {10.1093/mnras/stab2247},
  archiveprefix = {arXiv},
  eprint = {2011.10627},
  primaryclass = {astro-ph.GA},
  adsurl = {https://ui.adsabs.harvard.edu/abs/2021MNRAS.507.1662M},
  url = {https://doi.org/10.1093/mnras/stab2247}
}

@article{Misner+64,
  author = "Misner, Charles W. and Sharp, David H.",
  title = "{Relativistic equations for adiabatic, spherically symmetric gravitational collapse}",
  doi = "10.1103/PhysRev.136.B571",
  journal = "Phys. Rev.",
  volume = "136",
  pages = "B571--B576",
  year = "1964",
  url = {https://doi.org/10.1103/PhysRev.136.B571}
}

@article{Murray+13,
  author = "Murray, Steven and Power, Chris and Robotham, A. S. G.",
  title = "{HMFcalc: An online tool for calculating dark matter halo mass functions}",
  eprint = "1306.6721",
  archiveprefix = "arXiv",
  primaryclass = "astro-ph.CO",
  doi = "10.1016/j.ascom.2013.11.001",
  journal = "Astron. Comput.",
  volume = "3-4",
  pages = "23--34",
  year = "2013",
  url = {https://doi.org/10.1016/j.ascom.2013.11.001}
}

@article{Naidu+25,
  author = "Naidu, Rohan P. and others",
  title = "{A ``Black Hole Star'' Reveals the Remarkable Gas-Enshrouded Hearts of the Little Red Dots}",
  eprint = "2503.16596",
  archiveprefix = "arXiv",
  primaryclass = "astro-ph.GA",
  journal = "arXiv e-prints",
  year = "2025",
  url = {https://arxiv.org/abs/2503.16596}
}

@article{Navarro+96,
  author = "Navarro, Julio F. and Frenk, Carlos S. and White, Simon D. M.",
  title = "{The Structure of cold dark matter halos}",
  eprint = "astro-ph/9508025",
  archiveprefix = "arXiv",
  doi = "10.1086/177173",
  journal = "Astrophys. J.",
  volume = "462",
  pages = "563--575",
  year = "1996",
  url = {https://doi.org/10.1086/177173}
}

@article{Outmezguine+23,
  author = "Outmezguine, Nadav Joseph and Boddy, Kimberly K. and Gad-Nasr, Sophia and Kaplinghat, Manoj and Sagunski, Laura",
  title = "{Universal gravothermal evolution of isolated self-interacting dark matter halos for velocity-dependent cross-sections}",
  eprint = "2204.06568",
  archiveprefix = "arXiv",
  primaryclass = "astro-ph.GA",
  doi = "10.1093/mnras/stad1705",
  journal = "Mon. Not. Roy. Astron. Soc.",
  volume = "523",
  number = "3",
  pages = "4786--4800",
  year = "2023",
  url = {https://doi.org/10.1093/mnras/stad1705}
}

@article{Palubski+24,
  author = "Palubski, Igor and Slone, Oren and Kaplinghat, Manoj and Lisanti, Mariangela and Jiang, Fangzhou",
  title = "{Numerical challenges in modeling gravothermal collapse in Self-Interacting Dark Matter halos}",
  eprint = "2402.12452",
  archiveprefix = "arXiv",
  primaryclass = "astro-ph.CO",
  doi = "10.1088/1475-7516/2024/09/074",
  journal = "JCAP",
  volume = "09",
  pages = "074",
  year = "2024",
  url = {https://doi.org/10.1088/1475-7516/2024/09/074}
}

@article{Penrose+65,
  author = "Penrose, Roger",
  title = "{Gravitational collapse and space-time singularities}",
  doi = "10.1103/PhysRevLett.14.57",
  journal = "Phys. Rev. Lett.",
  volume = "14",
  pages = "57--59",
  year = "1965",
  url = {https://doi.org/10.1103/PhysRevLett.14.57}
}

@article{Perez-Gonzales+24,
  author = "P{\'e}rez-Gonz{\'a}lez, Pablo G. and others",
  title = "{What Is the Nature of Little Red Dots and What Is Not, MIRI SMILES Edition}",
  eprint = "2401.08782",
  archiveprefix = "arXiv",
  primaryclass = "astro-ph.GA",
  doi = "10.3847/1538-4357/ad38bb",
  journal = "Astrophys. J.",
  volume = "968",
  number = "1",
  pages = "4",
  year = "2024",
  url = {https://doi.org/10.3847/1538-4357/ad38bb}
}

@article{Pizzati+25,
  year = {2025},
  month = {4},
  title = {{‘Little red dots’ cannot reside in the same dark matter haloes as comparably luminous unobscured quasars}},
  author = {Pizzati, Elia and Hennawi, Joseph F and Schaye, Joop and Eilers, Anna-Christina and Huang, Jiamu and Schindler, Jan-Torge and Wang, Feige},
  journal = {Monthly Notices of the Royal Astronomical Society},
  issn = {0035-8711},
  doi = {10.1093/mnras/staf660},
  eprint = {2409.18208},
  pages = {2910--2925},
  number = {4},
  volume = {539},
  url = {https://doi.org/10.1093/mnras/staf660}
}

@article{Pollack+15,
  author = "Pollack, Jason and Spergel, David N. and Steinhardt, Paul J.",
  title = "{Supermassive Black Holes from Ultra-Strongly Self-Interacting Dark Matter}",
  eprint = "1501.00017",
  archiveprefix = "arXiv",
  primaryclass = "astro-ph.CO",
  reportnumber = "CALT-TH-2014-144",
  doi = "10.1088/0004-637X/804/2/131",
  journal = "Astrophys. J.",
  volume = "804",
  number = "2",
  pages = "131",
  year = "2015",
  url = {https://doi.org/10.1088/0004-637X/804/2/131}
}

@article{Reed+07,
  author = {Reed, Darren S. and Bower, Richard and Frenk, Carlos S. and Jenkins, Adrian and Theuns, Tom},
  title = {The halo mass function from the dark ages through the present day},
  journal = {Monthly Notices of the Royal Astronomical Society},
  volume = {374},
  number = {1},
  pages = {2-15},
  year = {2007},
  month = {01},
  issn = {0035-8711},
  doi = {10.1111/j.1365-2966.2006.11204.x},
  url = {https://doi.org/10.1111/j.1365-2966.2006.11204.x},
  eprint = {https://academic.oup.com/mnras/article-pdf/374/1/2/2835466/mnras0374-0002.pdf}
}

@article{Roberts+25,
  author = "Roberts, M. Grant and others",
  title = "{Early formation of supermassive black holes from the collapse of strongly self-interacting dark matter}",
  eprint = "2410.17480",
  archiveprefix = "arXiv",
  primaryclass = "astro-ph.GA",
  doi = "10.1088/1475-7516/2025/01/060",
  journal = "JCAP",
  volume = "01",
  pages = "060",
  year = "2025",
  url = {https://doi.org/10.1088/1475-7516/2025/01/060}
}

@article{Rocha+13,
  author = "Rocha, Miguel and Peter, Annika H. G. and Bullock, James S. and Kaplinghat, Manoj and Garrison-Kimmel, Shea and Onorbe, Jose and Moustakas, Leonidas A.",
  title = "{Cosmological Simulations with Self-Interacting Dark Matter I: Constant Density Cores and Substructure}",
  eprint = "1208.3025",
  archiveprefix = "arXiv",
  primaryclass = "astro-ph.CO",
  doi = "10.1093/mnras/sts514",
  journal = "Mon. Not. Roy. Astron. Soc.",
  volume = "430",
  pages = "81--104",
  year = "2013",
  url = {https://doi.org/10.1093/mnras/sts514}
}

@article{Roy+26,
  year = {2026},
  month = {6},
  title = {{Little Red Dots on FIRE: Exploring the formation and observational signatures of ultra-compact early galaxies}},
  author = {Roy, Niranjan Chandra and Anglés-Alcázar, Daniel and Cochrane, Rachel K and Richings, Alexander J and Mercedes-Feliz, Jonathan and Hayward, Christopher C and Faucher-Giguère, Claude-André and Lambrides, Erini and Feldmann, Robert and Oh, Boon Kiat and Marszewski, Andrew and Sun, Guochao and Davis, Kelcey and McKinney, Jed and Casey, Caitlin M and Díaz-Santos, Tanio and Brooks, Madisyn and Farrell, Grace},
  journal = {arXiv},
  doi = {10.48550/arXiv.2606.23683},
  eprint = {2606.23683},
  url = {https://arxiv.org/abs/2606.23683}
}

@article{Sabarish+25,
  author = {Sabarish, V. M. and Br{\"u}ggen, Marcus and Schmidt-Hoberg, Kai and Fischer, Moritz S.},
  title = "{Accretion of self-interacting dark matter onto supermassive black holes}",
  eprint = "2505.14779",
  archiveprefix = "arXiv",
  primaryclass = "astro-ph.CO",
  doi = "10.1051/0004-6361/202555586",
  journal = "Astron. Astrophys.",
  volume = "703",
  pages = "A142",
  year = "2025",
  url = {https://doi.org/10.1051/0004-6361/202555586}
}

@article{Shankar+04,
  year = {2004},
  month = {11},
  title = {{Supermassive black hole demography: the match between the local and accreted mass functions}},
  author = {Shankar, F. and Salucci, P. and Granato, G. L. and Zotti, G. De and Danese, L.},
  journal = {Monthly Notices of the Royal Astronomical Society},
  issn = {0035-8711},
  doi = {10.1111/j.1365-2966.2004.08261.x},
  eprint = {astro-ph/0405585},
  pages = {1020--1030},
  number = {4},
  volume = {354},
  url = {https://doi.org/10.1111/j.1365-2966.2004.08261.x}
}

@article{Shapiro+18,
  author = "Shapiro, Stuart L.",
  title = "{Star clusters, self-interacting dark matter halos, and black hole cusps: The fluid conduction model and its extension to general relativity}",
  eprint = "1809.02618",
  archiveprefix = "arXiv",
  primaryclass = "astro-ph.HE",
  doi = "10.1103/PhysRevD.98.023021",
  journal = "Phys. Rev. D",
  volume = "98",
  number = "2",
  pages = "023021",
  year = "2018",
  url = {https://doi.org/10.1103/PhysRevD.98.023021}
}

@article{Shen+26,
  year = {2026},
  month = {5},
  title = {{The Lumina Project: The Demographics of Active Galactic Nuclei from Quasars to Little Red Dots at $z=3$}},
  author = {Shen, Xuejian and Zier, Oliver and Smith, Aaron and Liu, Rongrong and Kannan, Rahul and Bulichi, Teodora-Elena and Koehler, Sonja M and Springel, Volker and Vogelsberger, Mark and Hernquist, Lars and Naidu, Rohan P and Graaff, Anna de and Pizzati, Elia and Alexander, David M and Ho, Luis C and Kokorev, Vasily and Leung, Gene and Eilers, Anna-Christina and Hickox, Ryan C},
  journal = {arXiv},
  doi = {10.48550/arXiv.2605.24112},
  eprint = {2605.24112},
  url = {https://arxiv.org/abs/2605.24112}
}

@article{Silverman+26,
  author = "Silverman, Maya and others",
  title = "{Mergers Matter: Gravothermal Collapse in Dwarf Halos with Self-Interacting Dark Matter}",
  eprint = "2606.02566",
  archiveprefix = "arXiv",
  primaryclass = "astro-ph.GA",
  reportnumber = "FERMILAB-PUB-26-0348-T",
  journal = "arXiv e-prints",
  year = "2026",
  url = {https://arxiv.org/abs/2606.02566}
}

@article{Spergel+00,
  author = "Spergel, David N. and Steinhardt, Paul J.",
  title = "{Observational evidence for selfinteracting cold dark matter}",
  eprint = "astro-ph/9909386",
  archiveprefix = "arXiv",
  doi = "10.1103/PhysRevLett.84.3760",
  journal = "Phys. Rev. Lett.",
  volume = "84",
  pages = "3760--3763",
  year = "2000",
  url = {https://doi.org/10.1103/PhysRevLett.84.3760}
}

@article{Taylor+25,
  author = "Taylor, Anthony J. and others",
  title = "{CAPERS-LRD-z9: A Gas-Enshrouded Little Red Dot Hosting a Broad-Line Active Galactic Nucleus at $z=9.288$}",
  eprint = "2505.04609",
  archiveprefix = "arXiv",
  primaryclass = "astro-ph.GA",
  doi = "10.3847/2041-8213/ade789",
  journal = "Astrophys. J. Lett.",
  volume = "989",
  number = "1",
  pages = "L7",
  year = "2025",
  url = {https://doi.org/10.3847/2041-8213/ade789}
}

@ARTICLE{Tulin+13,
  author = {{Tulin}, Sean and {Yu}, Hai-Bo and {Zurek}, Kathryn M.},
  title = "{Beyond collisionless dark matter: Particle physics dynamics for dark matter halo structure}",
  journal = {\prd},
  year = 2013,
  month = jun,
  volume = {87},
  number = {11},
  eid = {115007},
  pages = {115007},
  doi = {10.1103/PhysRevD.87.115007},
  archiveprefix = {arXiv},
  eprint = {1302.3898},
  primaryclass = {hep-ph},
  adsurl = {https://ui.adsabs.harvard.edu/abs/2013PhRvD..87k5007T},
  url = {https://doi.org/10.1103/PhysRevD.87.115007}
}

@article{Tulin+18,
  author = "Tulin, Sean and Yu, Hai-Bo",
  title = "{Dark Matter Self-interactions and Small Scale Structure}",
  eprint = "1705.02358",
  archiveprefix = "arXiv",
  primaryclass = "hep-ph",
  doi = "10.1016/j.physrep.2017.11.004",
  journal = "Phys. Rept.",
  volume = "730",
  pages = "1--57",
  year = "2018",
  url = {https://doi.org/10.1016/j.physrep.2017.11.004}
}

@article{Frank+25,
  author = "van den Bosch, Frank C. and Dattathri, Shashank",
  title = "{Dynamics in the Cores of Self-Interacting Dark Matter Halos: Reduced Stalling and Accelerated Core Collapse}",
  eprint = "2511.14912",
  archiveprefix = "arXiv",
  primaryclass = "astro-ph.GA",
  year = {2026},
  doi = {10.33232/001c.157701},
  journal = {The Open Journal of Astrophysics},
  volume = {9},
  month = {2},
  url = {https://doi.org/10.33232/001c.157701}
}

@article{Vogelsberger+12,
  author = "Vogelsberger, Mark and Zavala, Jesus and Loeb, Abraham",
  title = "{Subhaloes in Self-Interacting Galactic Dark Matter Haloes}",
  eprint = "1201.5892",
  archiveprefix = "arXiv",
  primaryclass = "astro-ph.CO",
  doi = "10.1111/j.1365-2966.2012.21182.x",
  journal = "Mon. Not. Roy. Astron. Soc.",
  volume = "423",
  pages = "3740",
  year = "2012",
  url = {https://doi.org/10.1111/j.1365-2966.2012.21182.x}
}

@article{Volonteri+10,
  author = "Volonteri, Marta",
  title = "{Formation of Supermassive Black Holes}",
  eprint = "1003.4404",
  archiveprefix = "arXiv",
  primaryclass = "astro-ph.CO",
  doi = "10.1007/s00159-010-0029-x",
  journal = "Astron. Astrophys. Rev.",
  volume = "18",
  pages = "279--315",
  year = "2010",
  url = {https://doi.org/10.1007/s00159-010-0029-x}
}

@article{Wang+26,
  author = "Wang, Zihao and Jiang, Fangzhou and Zheng, Haonan and Shen, Xuejian and Jia, Zixiang and Ho, Luis C. and Inayoshi, Kohei and Jiang, Linhua",
  title = "{Halo assembly bias in the early Universe: a clustering probe of the origin of the Little Red Dots}",
  eprint = "2603.15736",
  archiveprefix = "arXiv",
  primaryclass = "astro-ph.GA",
  journal = "arXiv e-prints",
  year = "2026",
  url = {https://arxiv.org/abs/2603.15736}
}

@article{Weibel+24,
  author = "Weibel, Andrea and others",
  title = "{Galaxy Build-up in the First 1.5 Gyr of Cosmic History: Insights from the Stellar Mass Function at $z\sim4$--$9$ from JWST NIRCam Observations}",
  eprint = "2403.08872",
  archiveprefix = "arXiv",
  primaryclass = "astro-ph.GA",
  doi = "10.1093/mnras/stae1891",
  journal = "Mon. Not. Roy. Astron. Soc.",
  volume = "533",
  number = "2",
  pages = "1808--1838",
  year = "2024",
  url = {https://doi.org/10.1093/mnras/stae1891}
}

@article{Yang+21,
  author = "Yang, Daneng and Yu, Hai-Bo",
  title = "{Self-interacting dark matter and small-scale gravitational lenses in galaxy clusters}",
  eprint = "2102.02375",
  archiveprefix = "arXiv",
  primaryclass = "astro-ph.GA",
  doi = "10.1103/PhysRevD.104.103031",
  journal = "Phys. Rev. D",
  volume = "104",
  number = "10",
  pages = "103031",
  year = "2021",
  url = {https://doi.org/10.1103/PhysRevD.104.103031}
}

@ARTICLE{Yang+22,
  author = {{Yang}, Daneng and {Yu}, Hai-Bo},
  title = "{Gravothermal evolution of dark matter halos with differential elastic scattering}",
  journal = {\jcap},
  year = 2022,
  month = sep,
  volume = {2022},
  number = {9},
  eid = {077},
  pages = {077},
  doi = {10.1088/1475-7516/2022/09/077},
  archiveprefix = {arXiv},
  eprint = {2205.03392},
  primaryclass = {astro-ph.CO},
  adsurl = {https://ui.adsabs.harvard.edu/abs/2022JCAP...09..077Y},
  url = {https://doi.org/10.1088/1475-7516/2022/09/077}
}

@article{Yu+26,
  author = "Yu, Hai-Bo",
  title = "{Core-Collapsed SIDM Halos as the Common Origin of Dense Perturbers in Lenses, Streams, and Satellites}",
  eprint = "2510.11006",
  archiveprefix = "arXiv",
  primaryclass = "astro-ph.GA",
  doi = "10.1103/txxx-97ln",
  journal = "Phys. Rev. Lett.",
  volume = "136",
  number = "14",
  pages = "141001",
  year = "2026",
  url = {https://doi.org/10.1103/txxx-97ln}
}

@article{Zeng+25,
  author = "Zeng, Zhichao Carton and Peter, Annika H. G. and Du, Xiaolong and Yang, Shengqi and Benson, Andrew and Cyr-Racine, Francis-Yan and Jiang, Fangzhou and Mace, Charlie and Metcalf, R. Benton",
  title = "{Evolution and properties of self-interacting dark matter subhalos until core collapse}",
  eprint = "2310.09910",
  archiveprefix = "arXiv",
  primaryclass = "astro-ph.GA",
  doi = "10.1103/PhysRevD.111.063001",
  journal = "Phys. Rev. D",
  volume = "111",
  number = "6",
  pages = "063001",
  year = "2025",
  url = {https://doi.org/10.1103/PhysRevD.111.063001}
}

@ARTICLE{Zhang+25,
  author = {{Zhang}, Ziwen and {Chen}, Yangyao and {Rong}, Yu and {Wang}, Huiyuan and {Mo}, Houjun and {Luo}, Xiong and {Li}, Hao},
  title = "{Unexpected clustering pattern in dwarf galaxies challenges formation models}",
  journal = {\nat},
  year = 2025,
  month = jun,
  volume = {642},
  number = {8066},
  pages = {47-52},
  doi = {10.1038/s41586-025-08965-5},
  archiveprefix = {arXiv},
  eprint = {2504.03305},
  primaryclass = {astro-ph.CO},
  adsurl = {https://ui.adsabs.harvard.edu/abs/2025Natur.642...47Z},
  url = {https://doi.org/10.1038/s41586-025-08965-5}
}

@article{Zhuang+25,
  author = "Zhuang, Ming-Yang and others",
  title = "{NEXUS: A Spectroscopic Census of Broad-Line AGNs and Little Red Dots at $3\lesssim z\lesssim6$}",
  eprint = "2505.20393",
  archiveprefix = "arXiv",
  primaryclass = "astro-ph.GA",
  journal = {Astrophys. J.},
  year = {2026},
  doi = {10.3847/1538-4357/ae3612},
  volume = {999},
  number = {1},
  pages = {31},
  month = {3},
  url = {https://doi.org/10.3847/1538-4357/ae3612}
}

\end{document}